\documentclass[prd,showpacs,preprintnumbers,amsmath,amssymb]{revtex4}
\usepackage{cancel}
\usepackage{booktabs}
\usepackage{multirow}
\usepackage{slashed}
\usepackage{graphicx}
 \usepackage{color}
\usepackage{epstopdf}
\begin{document}

\title{The bottomed strange molecules with isospin 0}

\author{Zhi-Feng Sun$^{1}$}\email{sunzf@lzu.edu.cn}
\author{Ju-Jun Xie$^{2}$}\email{xiejujun@impcas.ac.cn}
\author{E. Oset$^{3}$}\email{Eulogio.Oset@ific.uv.es}

\affiliation{$^1$School of Physical Science and Technology, Lanzhou University, Lanzhou 730000, China\\
$^2$Institute of Modern Physics, Chinese Academy of Sciences, Lanzhou 730000, China\\
$^3$Departamento de Fsica Terica and IFIC, Centro Mixto Universidad
de Valencia-CSIC, Institutos de Investigac$\acute{i}$on de Paterna,
Aptdo. 22085, 46071 Valencia, Spain}

\begin{abstract}
Using the local hidden gauge approach, we study the possibility of
the existence of bottomed strange molecular states with isospin 0.
We find three bound states with spin-parity $0^+$, $1^+$ and $2^+$
generated by the $\bar{K}^*B^*$ and $\omega B_s^*$ interaction,
among which the state with spin 2 can be identified as
$B_{s2}^*(5840)$. In addition, we also study the $\bar{K}^*B$ and
$\omega B_s$ interaction and find a bound state which can be
associated to $B_{s1}(5830)$. Besides, the $\bar{K}B^*$ and $\eta
B_s^*$ and $\bar{K}B$ and $\eta B_s$ systems are studied, and two
bound states are predicted. We expect that further experiments can
confirm our predictions.
\end{abstract}
\pacs{14.40.Be, 13.25.Jx, 12.38.Lg}
\date{\today}
\maketitle

\section{Introduction}\label{sec1}
The local hidden gauge symmetry was introduced in Refs.
\cite{Bando:1984ej,Bando:1987br,Meissner:1987ge,Harada:2003jx} which
regards vector mesons as the gauge bosons and pseudoscalar mesons as
the Goldstone bosons. Considering this symmetry together with the
global chiral symmetry, one can construct the Lagrangian describing
interactions involving vector and pseudoscalar mesons. On the other
hand, the Bethe-Salpeter equation is a powerful tool to deal with
nonperturbative physics restoring two body unitarity in coupled
channels. The theory incorporating the above two points has been
instrumental in explaining many properties of hadronic resonances.
In Ref. \cite{Molina:2008jw}, the $f_0(1370)$ and $f_2(1270)$ are
explained as resonances generated from $\rho\rho$ interaction.
Later, in Ref. \cite{Geng:2008gx} the work of \cite{Molina:2008jw}
was extended to SU(3), and five of the generated states can be
identified with the observed $f_0(1370)$, $f_2(1270)$, $f_0(1710)$,
$f_2^\prime(1525)$, $K_2^*(1430)$. In the spin 1 sector, a resonance
was also found in Ref. \cite{Geng:2008gx} with mass and width around
1800 and 80 MeV, respectively. This state, $h_1(1800)$, is
dynamically generated from the $K^*\bar{K}^*$ interaction, and it
was investigated in the $J/\psi\to \eta K^{*0}\bar{K}^{*0}$ in Ref.
\cite{Xie:2013ula} and in the $\eta_c\to \phi K^*\bar{K}^*$ in Ref.
\cite{Ren:2014ixa}. In Ref. \cite{Molina:2009eb}, the authors
studied the interactions of $\rho$, $\omega$ and $D^*$, and three
states with spin $J=0, 1, 2$ were predicted, among which the second
and the third ones are identified with $D^*(2640)$ and
$D^*_2(2460)$, respectively. The third state predicted, $D(2600)$,
was found later by \cite{delAmoSanchez:2010vq} and has been
reconfirmed \cite{Aaij:2013sza, Aaij:2016fma}. This work was
extended to the case of $\rho(\omega)B^*(B)$ interaction in Ref.
\cite{Soler:2015hna}, where $B_1(5721)$ and $B_2^*(5747)$ are
explained as $\rho (\omega) B^*$ and $\rho B$ molecules.

First evidence for at least one of the bottomed strange states was
found by the OPAL experiment \cite{Akers:1994fz}. Evidence for a
single state interpreted as $B_{s2}^*$ was seen by the Delphi
Collaboration \cite{Moch:2005ha}.
$B_{s2}^*(5840)$ was observed by both CDF and D0 in the $B^+K^-$
channel \cite{Mommsen:2006ai,Aaltonen:2007ah,Abazov:2007af}. In the
CDF experiment, there is another peak in the $B^+K^-$ invariant mass
spectrum corresponding to $B_{s1}(5830)$. However, $B_{s1}(5830)\to
B^+K^-$ is not allowed. The interpretation is that this peak comes
from the channel $B^{*+}K^-$ and $B^{*+}$ decays to $B^+\gamma$
where the photon is not detected. As a consequence, the peak is
shifted by $B^*-B$ mass difference due to the missing momentum of
the photon. Recently, LHCb first measured the mass and width of
$B_{s2}^*(5840)$ in the $B^{*+}K^-$ channel. Besides, the ratio
$\frac{B_{s2}^*(5840)\to B^{*+}K^-}{B_{s2}^*(5840)\to B^{+}K^-}$ was
also measured and the decay of $B_{s1}(5830)\to B^{*+}K^-$ was
observed as well \cite{Aaij:2012uva}.

In this work, we extrapolate the local hidden gauge approach to the
systems containing bottomed and strange quarks. The paper is
organized as follows. After this introduction, in section II we will
show the local hidden gauge Lagrangian, from which the potentials
are obtained. And then we construct the $T$ matrix by solving the
Bethe-Salpeter equation. In section III, the results are given.
Finally, we make a short summary.

\section{Formalism}
\subsection{Lagrangian}
In order to describe the interaction of bottomed and strange mesons,
we need to use the local hidden gauge approach, under which vector
mesons are treated as gauge bosons. The covariant derivative is
defined as
\begin{eqnarray}
D_\mu \xi_{L,R}=\partial_\mu \xi_{L,R}-iV_\mu \xi_{L,R},
\end{eqnarray}
and the gauge field strength as
\begin{eqnarray}
V_{\mu\nu}=\partial_\mu V_\nu-\partial_\nu V_\mu-ig[V_\mu,V_\nu].
\end{eqnarray}
Here, $g$ is given by $ g=\frac{m_V}{2f_{\pi}}$ with the pion decay
constant $f_\pi=93$ MeV, and $m_V$ the mass of vector mesons.
$\xi_{L,R}$ is defined as
\begin{eqnarray}
\xi_{L}&=&e^{i\sigma/f_{\sigma}}e^{-i\frac{1}{\sqrt{2}}P/f_{\pi}},\\
\xi_{R}&=&e^{i\sigma/f_{\sigma}}e^{i\frac{1}{\sqrt{2}}P/f_{\pi}}.
\end{eqnarray}
In this paper, we take the unitary gauge, i.e., $\sigma=0$. In the
above equations, the matrices $V_\mu$ and $P$ have the following
form
\begin{small}
\begin{eqnarray}\label{matrix}
V_\mu&=&\left(\begin{array}{cccc}
\frac{\omega}{\sqrt{2}}+\frac{\rho^0}{\sqrt{2}}&\rho^+&K^{*+}&B^{*+}\\
\rho^-&\frac{\omega}{\sqrt{2}}-\frac{\rho^0}{\sqrt{2}}&K^{*0}&B^{*0}\\
K^{*-}&\bar{K}^{*0}&\phi &B_s^{*0}\\
B^{*-}&\bar{B}^{*0}&\bar{B}_s^{*0}&\Upsilon
\end{array}\right)_\mu,\nonumber\\
P&=&\left(\begin{array}{cccc}
\frac{\eta}{\sqrt{3}}+\frac{\eta^\prime}{\sqrt{6}}+\frac{\pi^0}{\sqrt{2}}&\pi^+&K^{+}&B^{+}\\
\pi^-&\frac{\eta}{\sqrt{3}}+\frac{\eta^\prime}{\sqrt{6}}-\frac{\pi^0}{\sqrt{2}}&K^{0}&B^{0}\\
K^{-}&\bar{K}^{0}&-\frac{\eta}{\sqrt{3}}+\sqrt{\frac{2}{3}}\eta^\prime &B_s^{0}\\
B^{-}&\bar{B}^{0}&\bar{B}_s^{0}&\eta_b
\end{array}\right).
\end{eqnarray}
\end{small} After defining the blocks
\begin{eqnarray}
\hat{\alpha}_{\perp\mu}=\frac{1}{2i}\left(D_\mu \xi_R \cdot
\xi^\dag_R-D_\mu \xi_L \cdot \xi^\dag_L\right),\nonumber\\
\hat{\alpha}_{\parallel\mu}=\frac{1}{2i}\left(D_\mu \xi_R \cdot
\xi^\dag_R+D_\mu \xi_L \cdot \xi^\dag_L\right),
\end{eqnarray}
one can construct the Lagrangian \cite{Harada:2003jx}
\begin{eqnarray}\label{Lagrangian}
\mathcal{L}&=&\mathcal{L}_A+a\mathcal{L}_V+\mathcal{L}_{III},
\end{eqnarray}
where
\begin{eqnarray}
\mathcal{L}_A&=&f_\pi^2\langle
\hat{\alpha}_{\perp\mu}\hat{\alpha}_{\perp}^\mu\rangle,\nonumber\\
a\mathcal{L}_V&=&f_\sigma^2\langle
\hat{\alpha}_{\parallel\mu}\hat{\alpha}_{\parallel}^\mu\rangle,\nonumber\\
\mathcal{L}_{III}&=&-\frac{1}{4}\langle V_{\mu\nu}V^{\mu\nu}\rangle,
\end{eqnarray}
with $f_\sigma^2=af_\pi^2$, and we take $a=2$ as in Ref.
\cite{Harada:2003jx}.

After expanding the Lagrangians in Eq. \eqref{Lagrangian}, we get
the terms needed in our calculation, i.e., three vector vertex
\begin{eqnarray}\label{LVVV}
\mathcal{L}_{VVV}=ig\langle (\partial_\mu V_\nu-\partial_\nu
V_\mu)V^\mu V^\nu\rangle,
\end{eqnarray}
four vector vertex
\begin{eqnarray}\label{LVVVV}
\mathcal{L}_{VVVV}=\frac{g^2}{2}\langle V_\mu V_\nu V^\mu
V^\nu-V_\nu V_\mu V^\mu V^\nu\rangle,
\end{eqnarray}
four pseudoscalar vertex
\begin{eqnarray}\label{LPPPP}
\mathcal{L}_{PPPP}=-\frac{1}{24f_\pi^2}\langle [P,\partial_\mu
P][P,\partial^\mu P]\rangle
\end{eqnarray}
and vector pseudoscalar pseudoscalar vertex
\begin{eqnarray}\label{LVPP}
\mathcal{L}_{VPP}=-ig\langle V_\mu[P,\partial^\mu P]\rangle.
\end{eqnarray}
Note that there is no $VVPP$ contact term under the hidden local
symmetry. Moreover, since the VVP interaction is anomalous with a
comparatively small contribution, we do not take it into account. In
this work, we will study the interaction between bottom and strange
mesons, so we extend the SU(3) flavor symmetry to SU(4). Next we
change the form of the three vector Lagrangian in Eq. (\ref{LVVV})
through some short calculations
\begin{eqnarray}\label{LVVVp}
\mathcal{L}&=&ig\langle (\partial_\mu V_\nu-\partial_\nu V_\mu)V^\mu
V^\nu\rangle\nonumber\\
&=&ig\langle \partial_\mu V_\nu V^\mu V^\nu-\partial_\nu V_\mu V^\mu
V^\nu\rangle\nonumber\\
&=&ig\langle \partial_\nu V_\mu V^\nu V^\mu-\partial_\nu V_\mu V^\mu
V^\nu\rangle\nonumber\\
&=&ig\langle V^\mu\partial_\nu V_\mu V^\nu -\partial_\nu V_\mu V^\mu
V^\nu\rangle\nonumber\\
&=&ig\langle [V_\mu ,\partial_\nu V^\mu ] V^\nu\rangle
\nonumber\\
&=&ig\langle V_\mu[V_\nu ,\partial^\mu V^\nu ] \rangle,
\end{eqnarray}
from which we see that this Lagrangian has a similar form as that in
Eq. (\ref{LVPP}) except for the minus sign.

As noted in \cite{Oset:2009vf}, for small three momenta of the
vector mesons compared to their mass, the $\epsilon^0$ component of
the external vectors can be neglected. Then $V_\mu$ in the last of
Eq. \eqref{LVVVp} should be $V_i$ (i=1, 2, 3) if it corresponds to
an external vector, but then $\partial^i$ will give a three momentum
of this vector, which one is neglecting. Hence $V_\mu$ cannot
correspond to the external vectors and is necessarily the exchanged
vector. The rest of the operator $V_\nu V^\nu$ gives rise to
$\epsilon_\mu \epsilon^\mu \to -\vec{\epsilon}\cdot \vec{\epsilon}$
and the last of the Eq. \eqref{LVVVp} is equivalent to Eq.
\eqref{LVPP} including the sign.

It should be noted that the local hidden gauge approach is
constructed within SU(2) or SU(3)
\cite{Ecker:1989yg,Nagahiro:2008cv}. In the heavy quark sector one
cannot invoke heavy mesons as Goldstone bosons. Yet, the extension
to the heavy quark sector is possible because the dominant terms of
the interaction correspond to the exchange of light vectors, $\rho,
\omega, \phi$ and the heavy quarks of the hadrons are just
spectators. In this case it is possible to make a mapping of the
interaction in the heavy light hadron sector to the one in the heavy
hadron sector. For practical purposes one can use the local hidden
gauge Lagrangians extrapolated to SU(4) as in Eq. \eqref{matrix},
since for the exchange of light vectors one is only making use of
the relevant SU(3) subgroup. Discussion on this issue and the proof
of this property can be seen in section II of \cite{Sakai:2017avl}
and section II and Appendix of \cite{Liang:2017ejq}.

\subsection{$B^*$ and $\bar{K}^*$ interaction}
\begin{figure*}
\begin{tabular}{cccccc}
\includegraphics[scale=0.22]{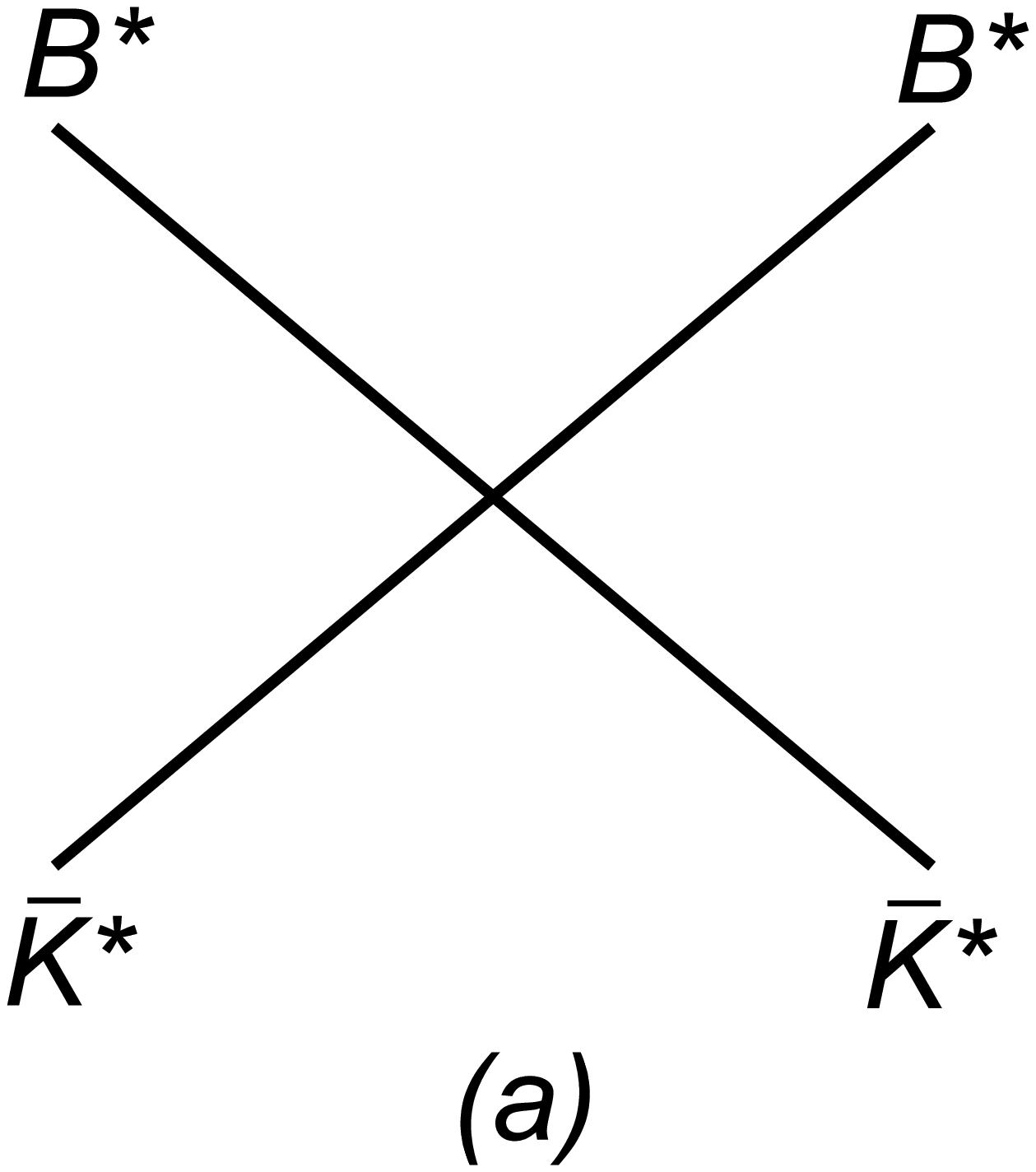}\quad\quad
\includegraphics[scale=0.22]{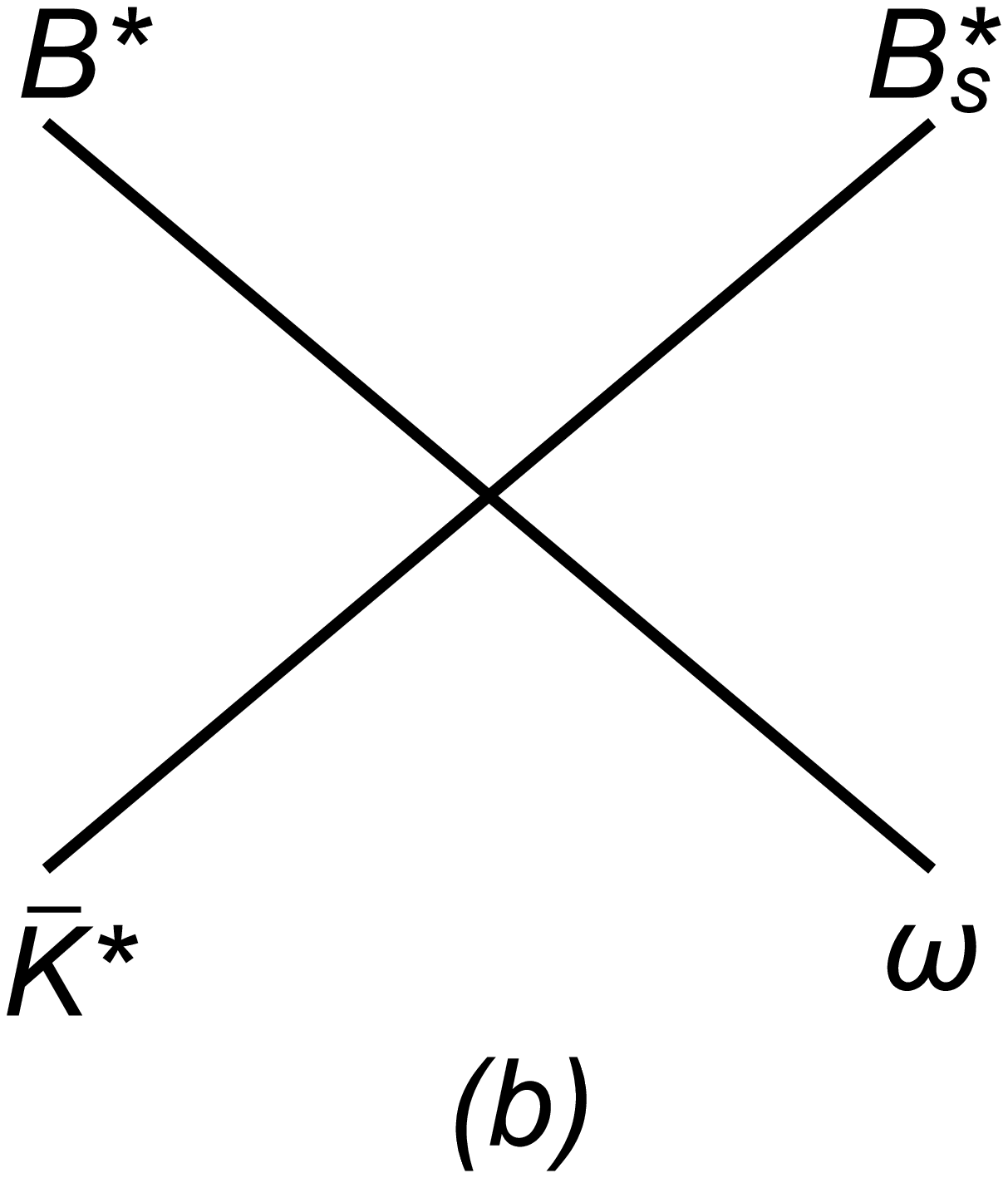}\quad\quad
\includegraphics[scale=0.22]{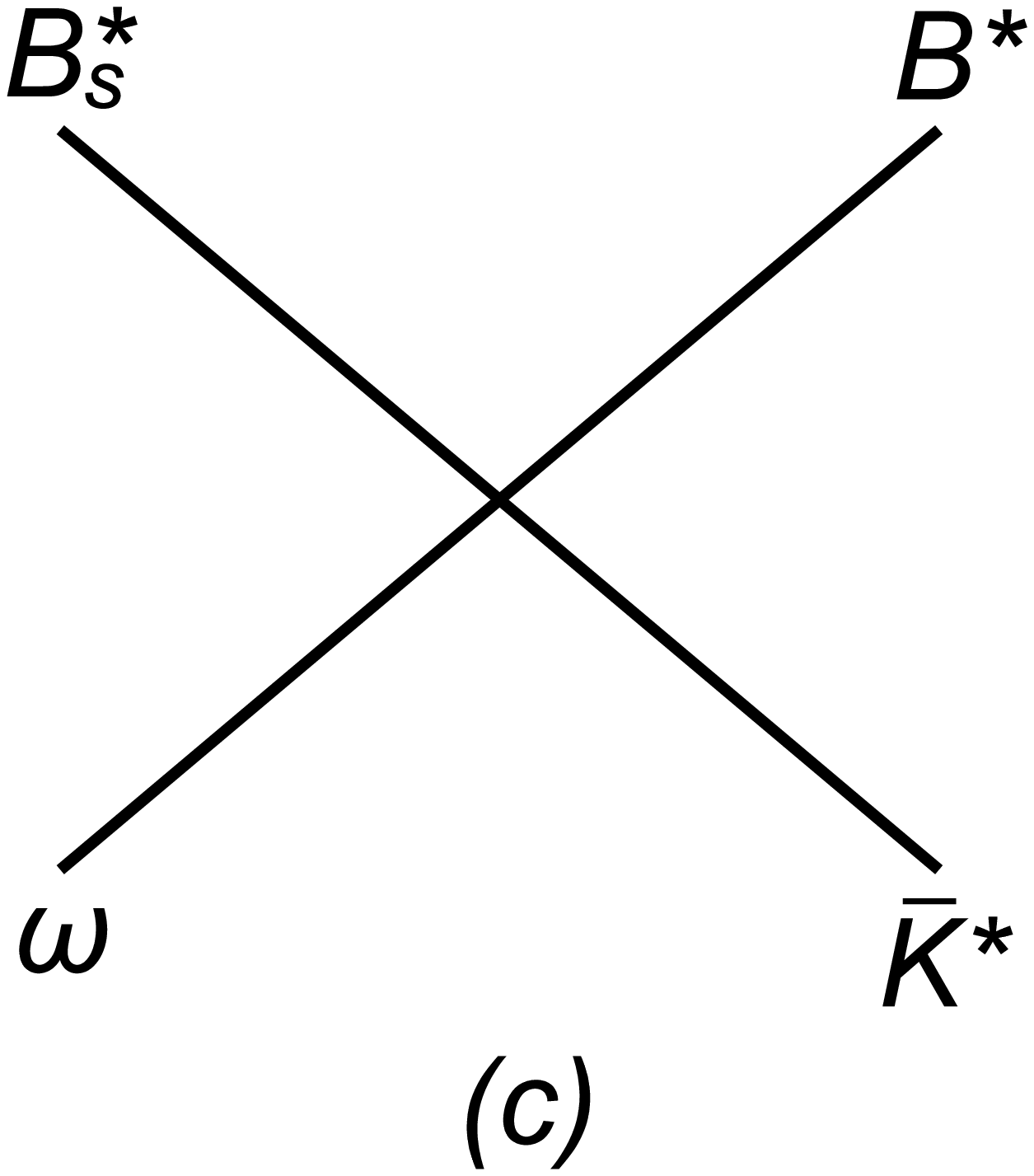}\quad\quad
\includegraphics[scale=0.22]{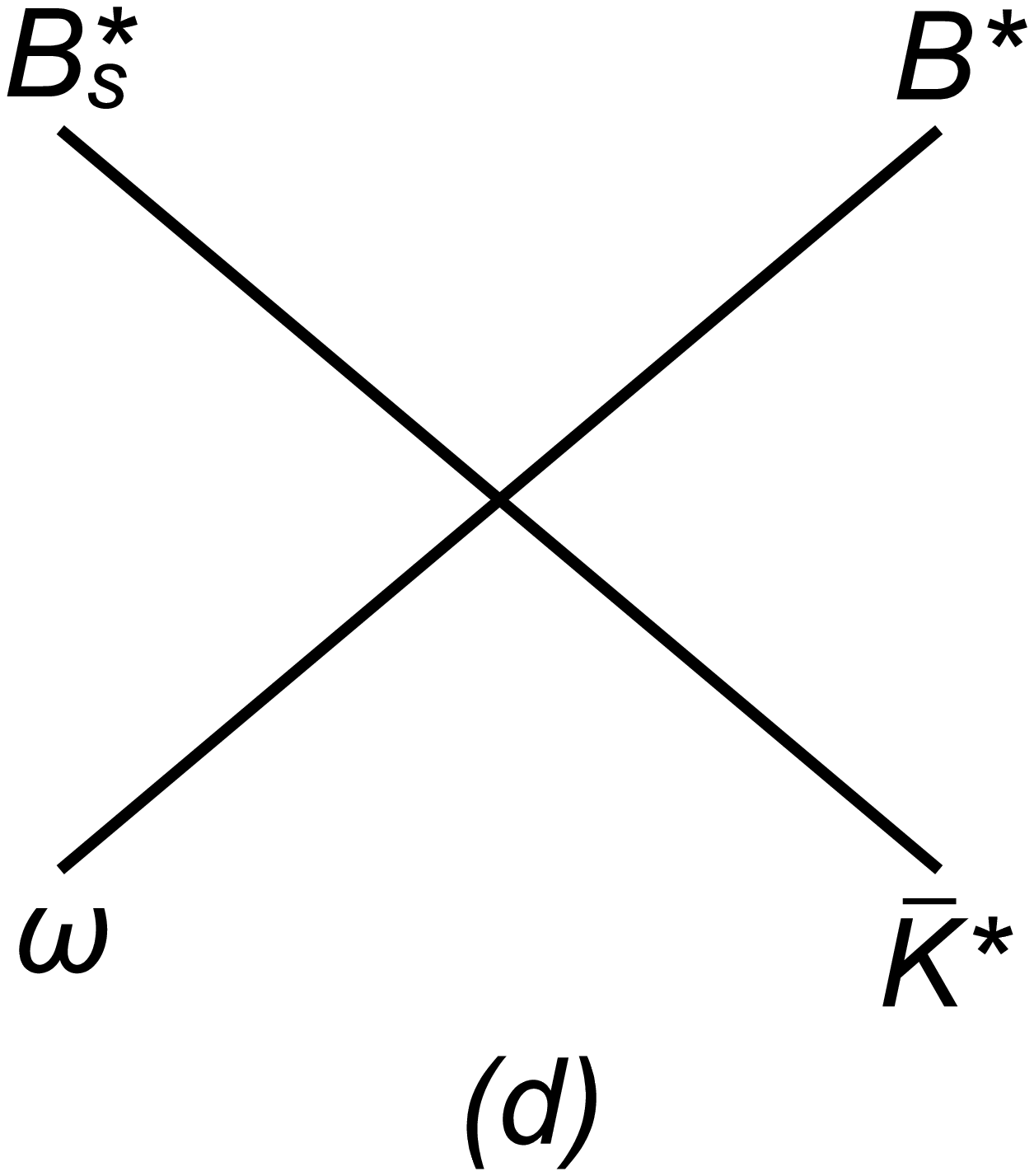}\\
\\
\\
\includegraphics[scale=0.22]{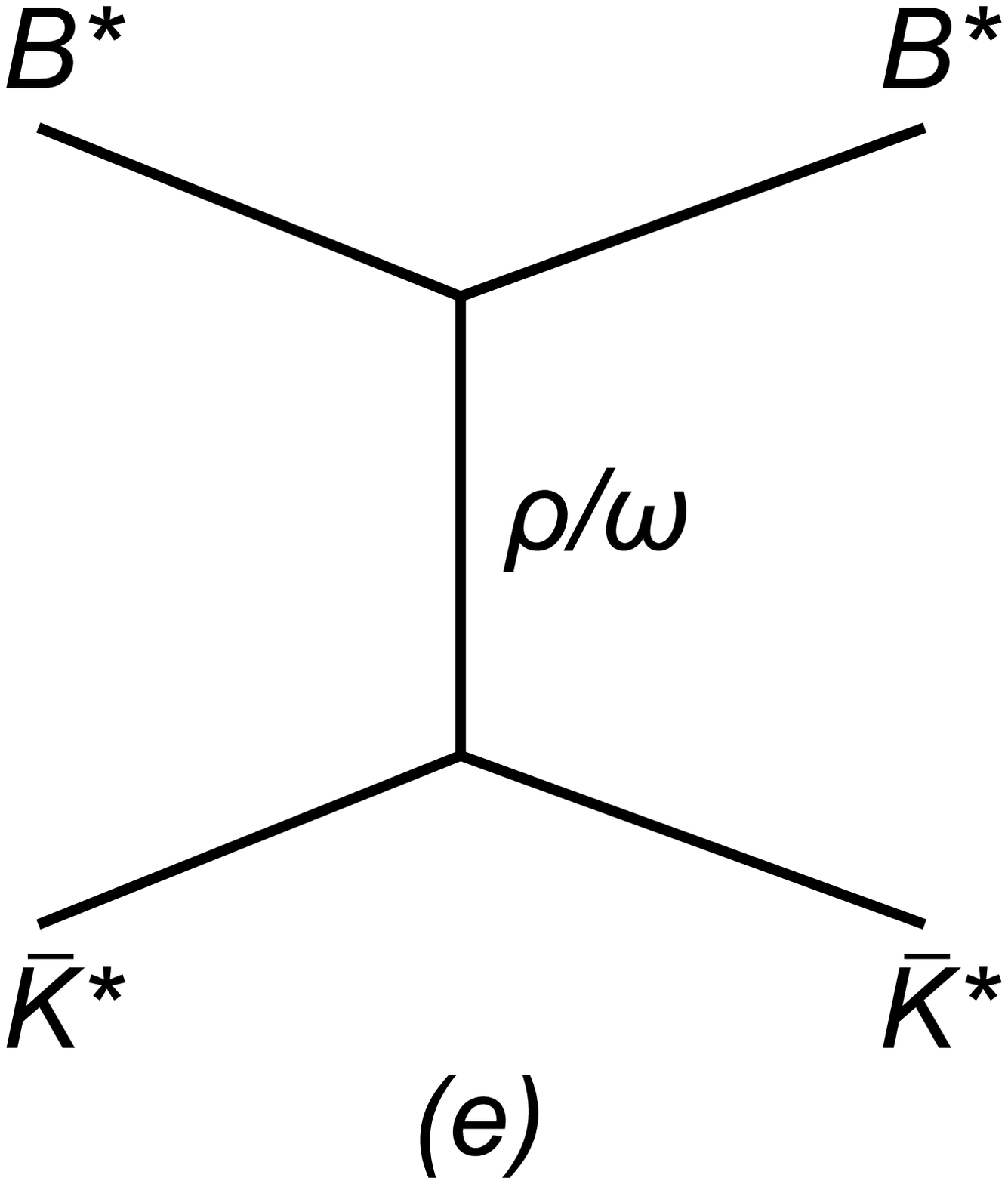}\quad\quad
\includegraphics[scale=0.22]{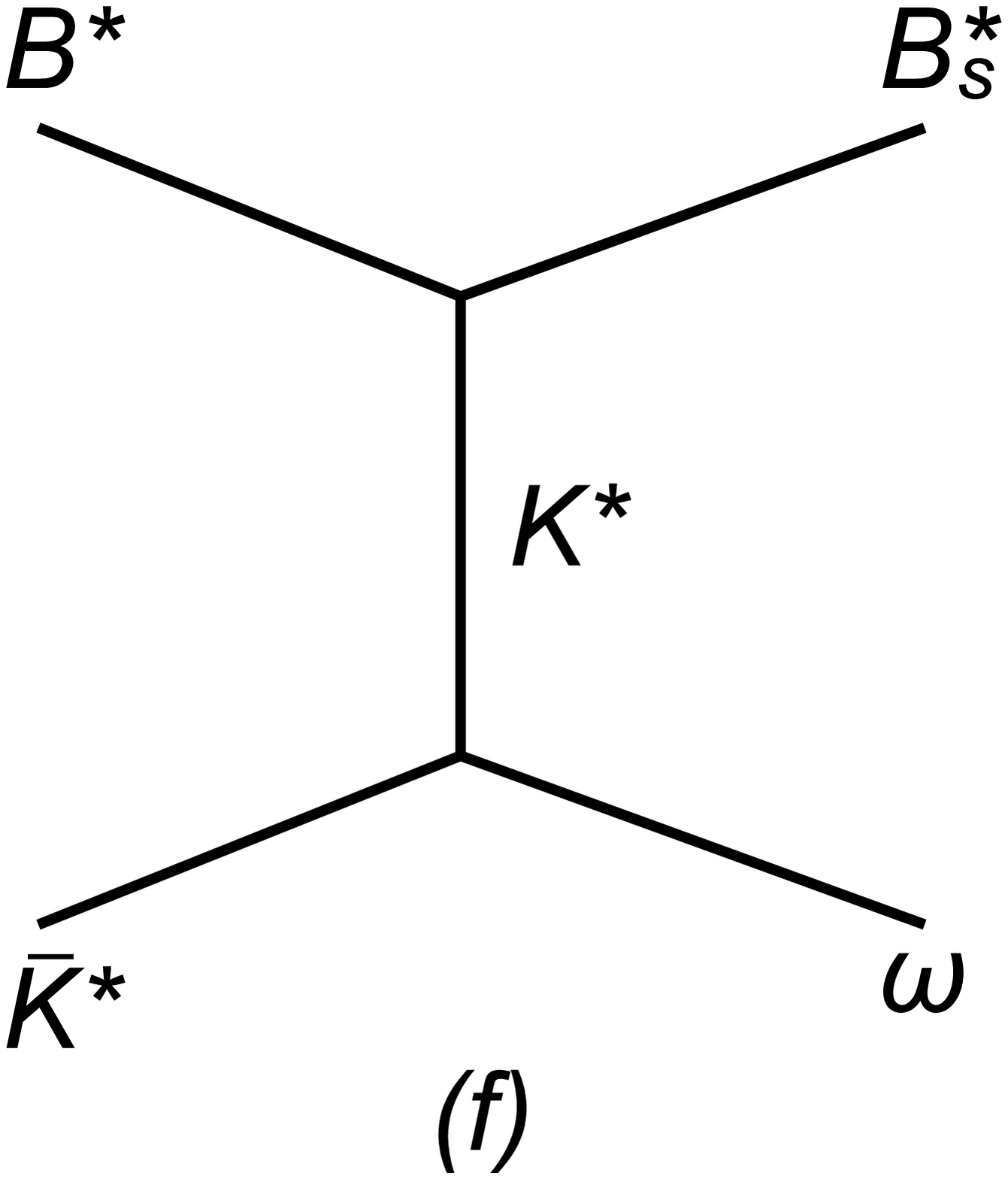}\quad\quad
\includegraphics[scale=0.22]{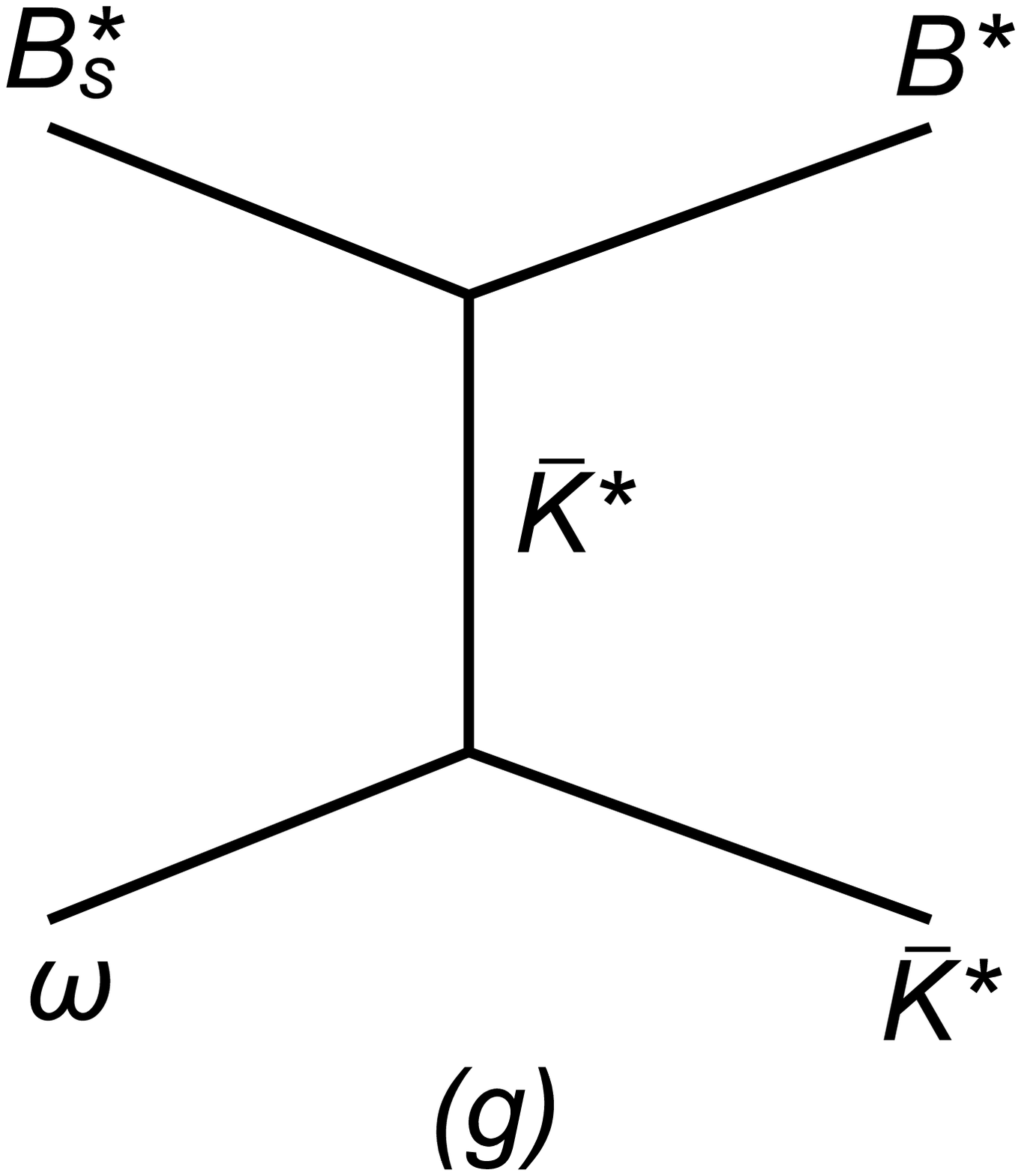}
\end{tabular}
\caption{Feynman diagrams describing $\bar{K}^*B^*$ and $\omega
B_s^*$ interaction.\label{fig1}}
\end{figure*}

The interaction terms of $\bar{K}^*B^*$ and $\omega B_s^*$ are
depicted by the diagrams in Fig. \ref{fig1}, including contact terms
and t-channel diagrams. Here, we neglect the bottomed-meson-exchange
diagrams, which have a much smaller contribution due to the heavy
mass of bottomed mesons. Besides, the amplitude of $\omega B_s^*\to
\omega B_s^*$ is zero, because of the OZI (Okubo-Zweig-Iizuka) rule
\cite{Okubo:1963fa,Zweig:1964jf,Iizuka:1966fk}. Recalling the
isospin doublet $(K^{*+},K^{*0})$, $(\bar{K}^{*0},-K^{*-})$,
$(B^{*+},B^{*0})$, $(\bar{B}^{*0},-B^{*-})$, and the isospin triplet
$(-\rho^+,\rho^0,\rho^-)$, we have the flavor wave functions
\begin{eqnarray}
|\bar{K}^*B^*;I=0\rangle&=&\frac{K^{*-}B^{*+}+\bar{K}^{*0}B^{*0}}{\sqrt{2}},\\
| \omega B_s^*;I=0\rangle&=&\omega B_s^*.
\end{eqnarray}
Here the channel $\phi B_s^*$ is not considered, since its threshold
is much higher than the other two. With the structure of Eqs.
\eqref{LVPP} and \eqref{LVVVp}, all the amplitudes have the
structure of $(k_1+k_3)\cdot(k_2+k_4)\epsilon_{\mu
1}\epsilon^{\mu}_3\epsilon_{\nu 2}\epsilon^{\nu}_4$. After writing
the amplitudes using Feynman rules, we project the polarization
vector products into different spin states:
\begin{eqnarray}
\mathcal{P}(0)&=&\frac{1}{3}\epsilon_\mu \epsilon^\mu \epsilon_\nu
\epsilon^\nu,\\
\mathcal{P}(1)&=&\frac{1}{2}(\epsilon_\mu \epsilon_\nu \epsilon^\mu
\epsilon^\nu-\epsilon_\mu \epsilon_\nu
\epsilon^\nu \epsilon^\mu),\\
\mathcal{P}(2)&=&\frac{1}{2}(\epsilon_\mu \epsilon_\nu \epsilon^\mu
\epsilon^\nu+\epsilon_\mu \epsilon_\nu \epsilon^\nu
\epsilon^\mu)-\frac{1}{3}\epsilon_\mu \epsilon^\mu \epsilon_\nu
\epsilon^\nu
\end{eqnarray}
with the order of the $\epsilon$'s as $1,2,3,4$ for the reaction
$1+2\to 3+4$. Hence we get the amplitudes of different spins for
$\bar{K}^*B^*\to \bar{K}^*B^*$ with $I=0$ as follows:
\begin{eqnarray}
t_{cont}^{S=0}&=&4g^2,\\
t_{cont}^{S=1}&=&6g^2,\\
t_{cont}^{S=2}&=&-2g^2,\\
t_{ex}^{S=0,1,2}&=&-\frac{g^2}{2}\left(\frac{3}{m_\rho^2}+\frac{1}{m_\omega^2}\right)(s-u),
\end{eqnarray}
for $\bar{K}^*B^*\to \omega B^*_s$
\begin{eqnarray}
t_{cont}^{S=0}&=&-4g^2,\\
t_{cont}^{S=1}&=&0,\\
t_{cont}^{S=2}&=&2g^2,\\
t_{ex}^{S=0,1,2}&=&\frac{g^2}{m_{K^*}^2}(s-u).
\end{eqnarray}
In the above equations, the Mandelstam variables $s$ and $u$ are
defined as
\begin{eqnarray}
s&=&(k_1+k_2)^2,\\
u&=&(k_1-k_4)^2.
\end{eqnarray}

\subsection{$B$ and $\bar{K}^*$ and $B^*$ and $\bar{K}$ interactions}
\begin{figure*}
\begin{tabular}{cccccc}
\includegraphics[scale=0.22]{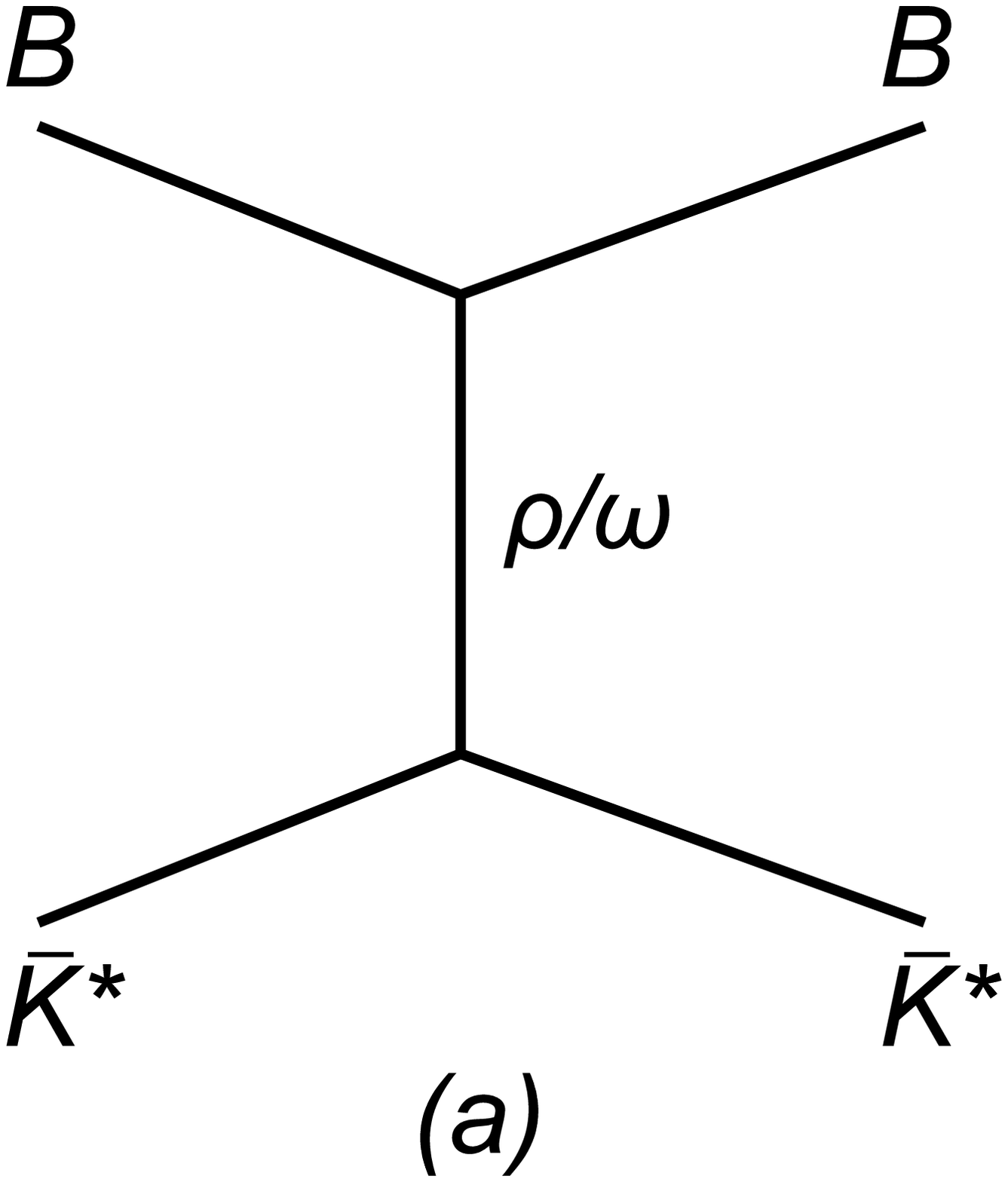}\quad\quad
\includegraphics[scale=0.22]{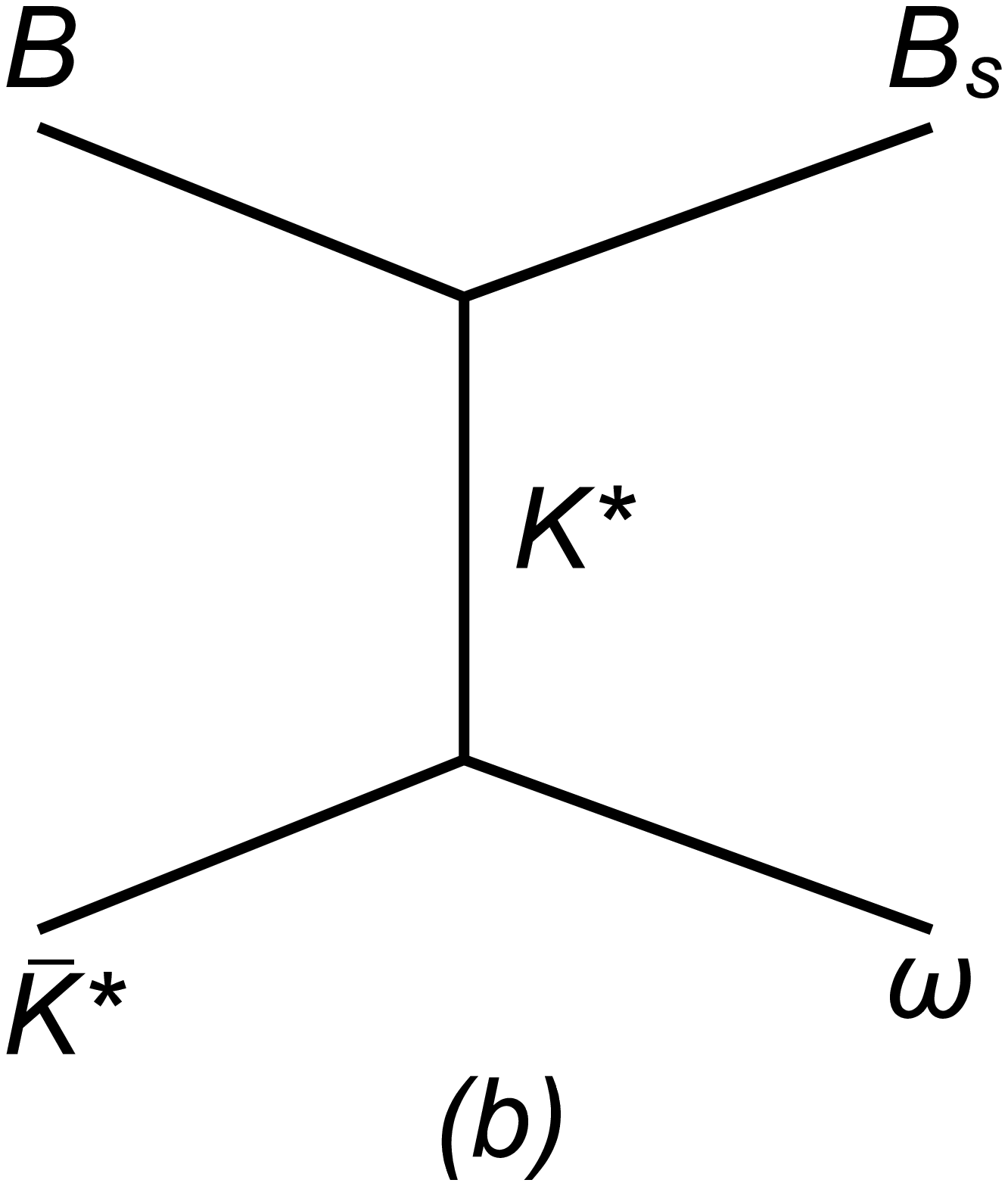}\quad\quad
\includegraphics[scale=0.22]{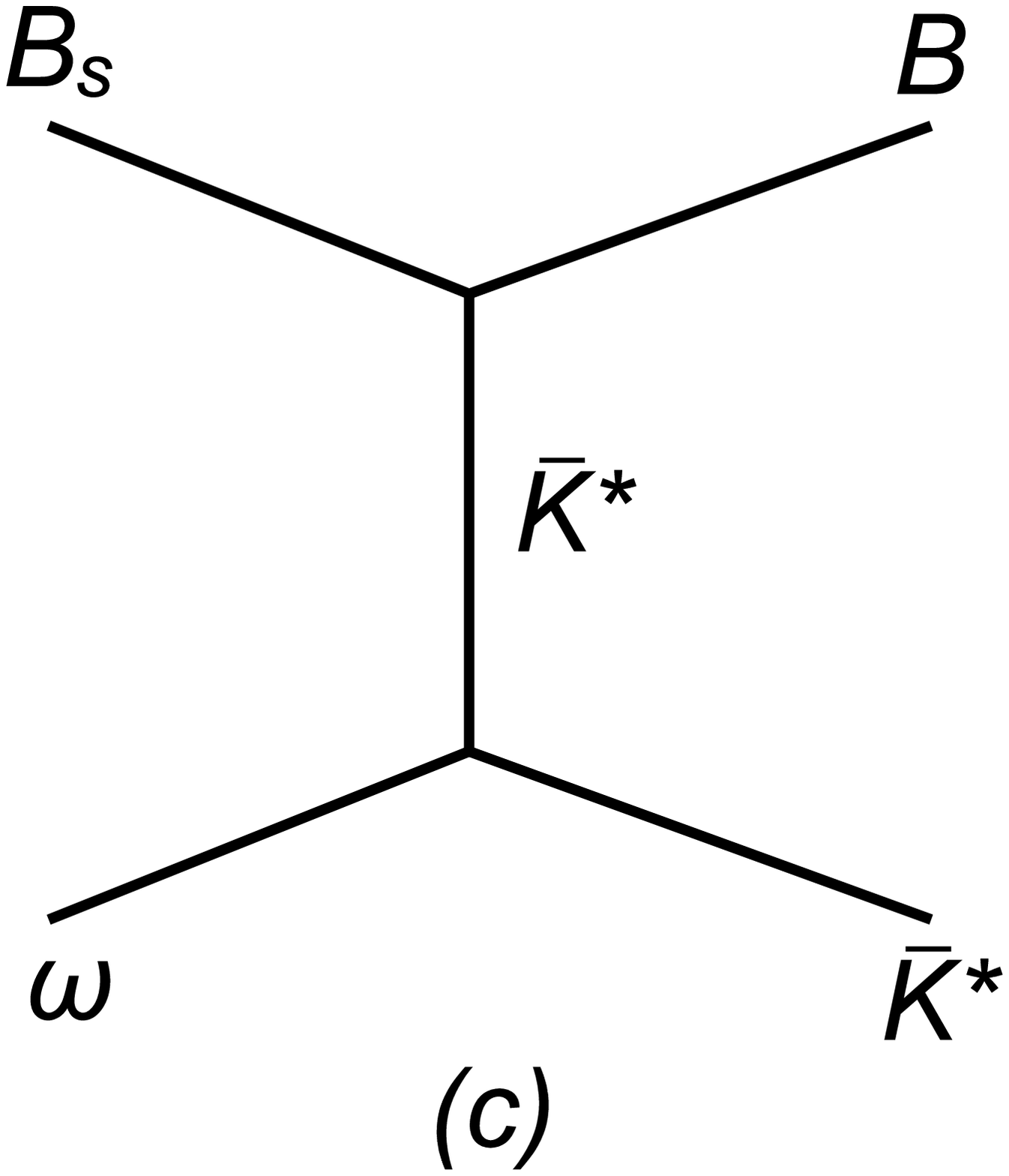}
\end{tabular}
\caption{Feynman diagrams describing $\bar{K}^*B$ and $\omega B_s$
interaction.\label{fig2}}
\end{figure*}

\begin{figure*}
\begin{tabular}{cccccc}
\includegraphics[scale=0.22]{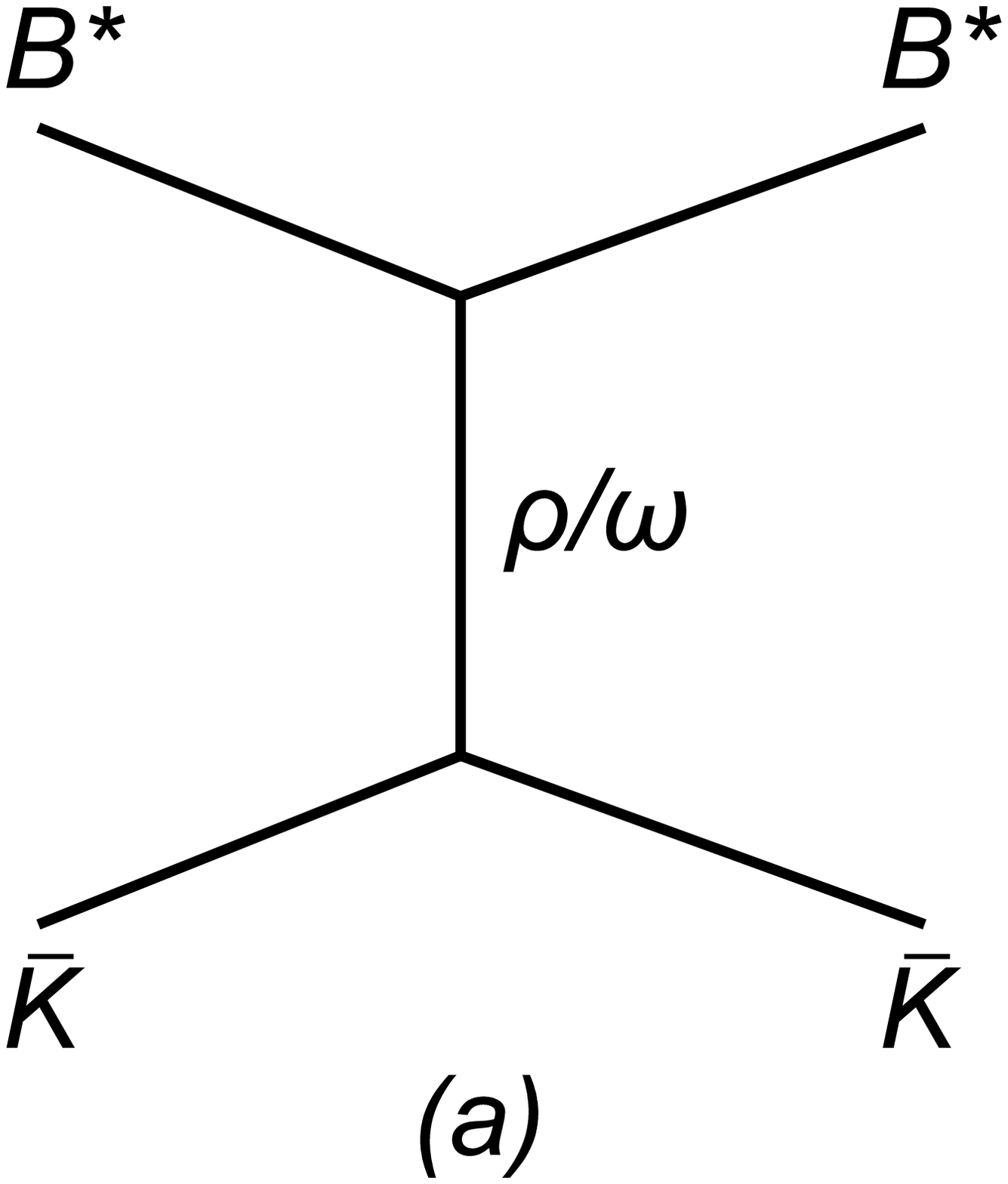}\quad\quad
\includegraphics[scale=0.22]{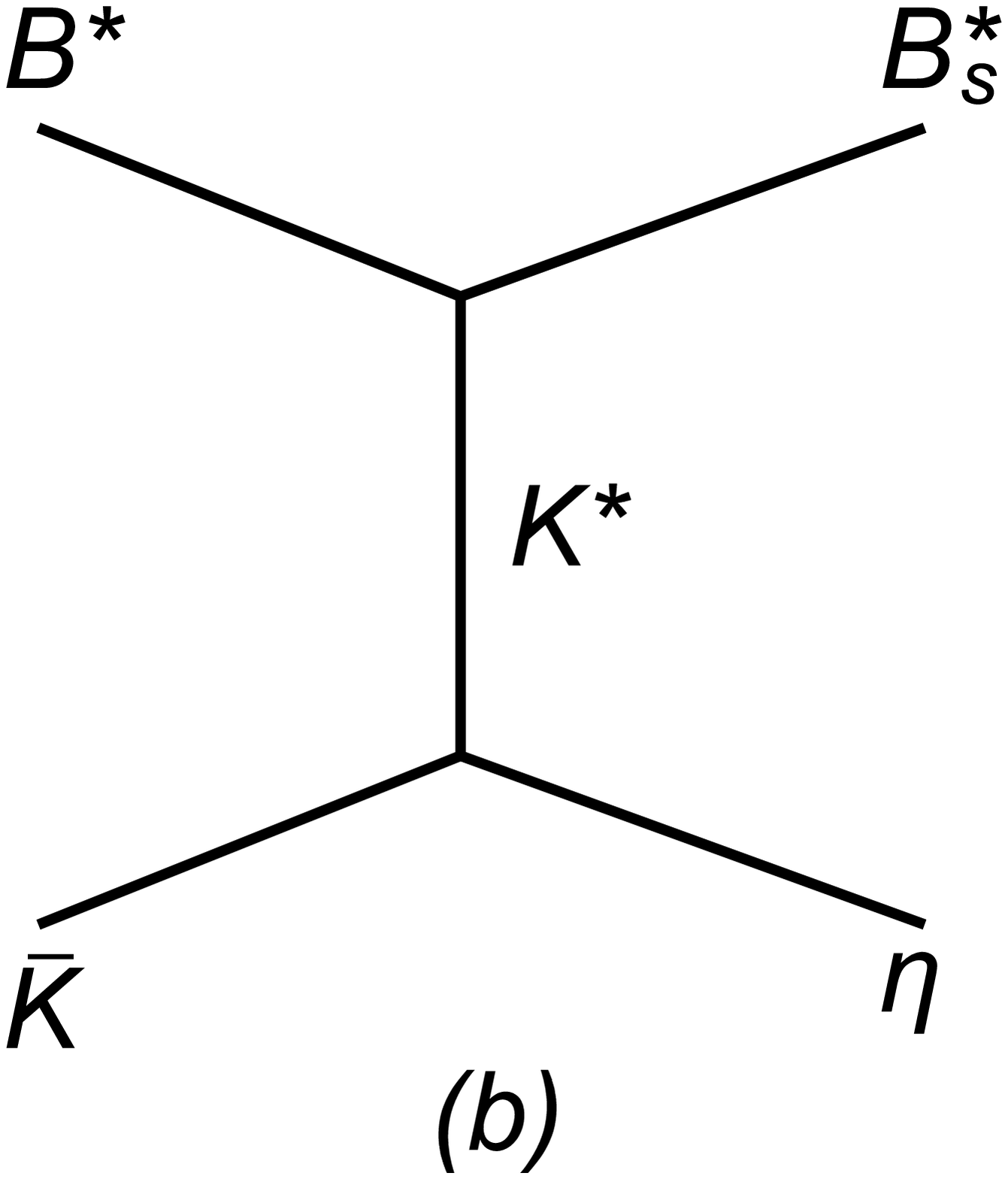}\quad\quad
\includegraphics[scale=0.22]{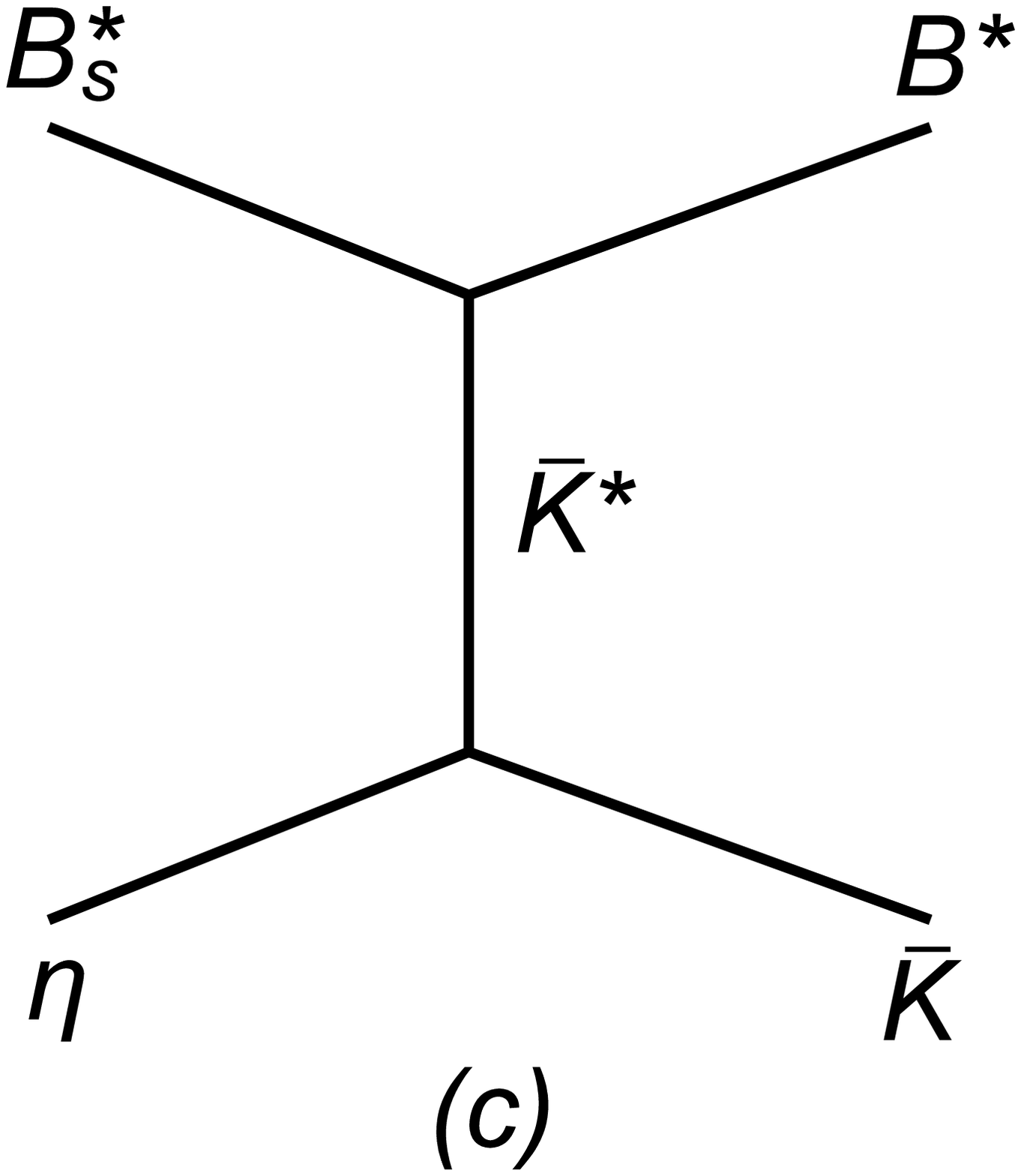}
\end{tabular}
\caption{Feynman diagrams describing $\bar{K}B^*$ and $\eta B_s^*$
interaction.\label{fig3}}
\end{figure*}

In Fig. \ref{fig2}, we show the diagrams for the $\bar{K}^*B$ and
$\omega B_s$ interaction. Note that under hidden local symmetry,
there is no contact term for vector pseudoscalar scattering. The
amplitude of $\omega B_s\to \omega B_s$ is zero, because of the OZI
rules.

For $\bar{K}^*B\to \bar{K}^*B$ in $I=0$ we need the exchange of
$\rho$ and $\omega$ and we obtain
\begin{eqnarray}
t_{ex}^{S=1}&=&-\frac{g^2}{2}\left(\frac{3}{m_\rho^2}+\frac{1}{m_\omega^2}\right)(s-u)
\end{eqnarray}
and for $\bar{K}^*B\to \omega B_s$
\begin{eqnarray}
t_{ex}^{S=1}&=&\frac{g^2}{m_{K^*}^2}(s-u).
\end{eqnarray}
Similarly, we can also get the amplitudes for the $\bar{K}B^*\to
\bar{K}B^*$ process in $I=0$ as follows
\begin{eqnarray}
t_{ex}^{S=1}&=&-\frac{g^2}{2}\left(\frac{3}{m_\rho^2}+\frac{1}{m_\omega^2}\right)(s-u).
\end{eqnarray}
However, according to the diagrams shown in Fig. \ref{fig3}, the
calculation for $\bar{K}B^*\to \eta B_s^*$ in $I=0$ is a little bit
different. Using Feynman rule and considering the flavor wave
function, we obtain
\begin{eqnarray}
t_{ex}^{S=1}&=&-\frac{2\sqrt{6}g^2}{3m_{K^*}^2}(s-u).
\end{eqnarray}

\subsection{$B$ and $\bar{K}$ interaction}
\begin{figure*}
\begin{tabular}{cccccc}
\includegraphics[scale=0.22]{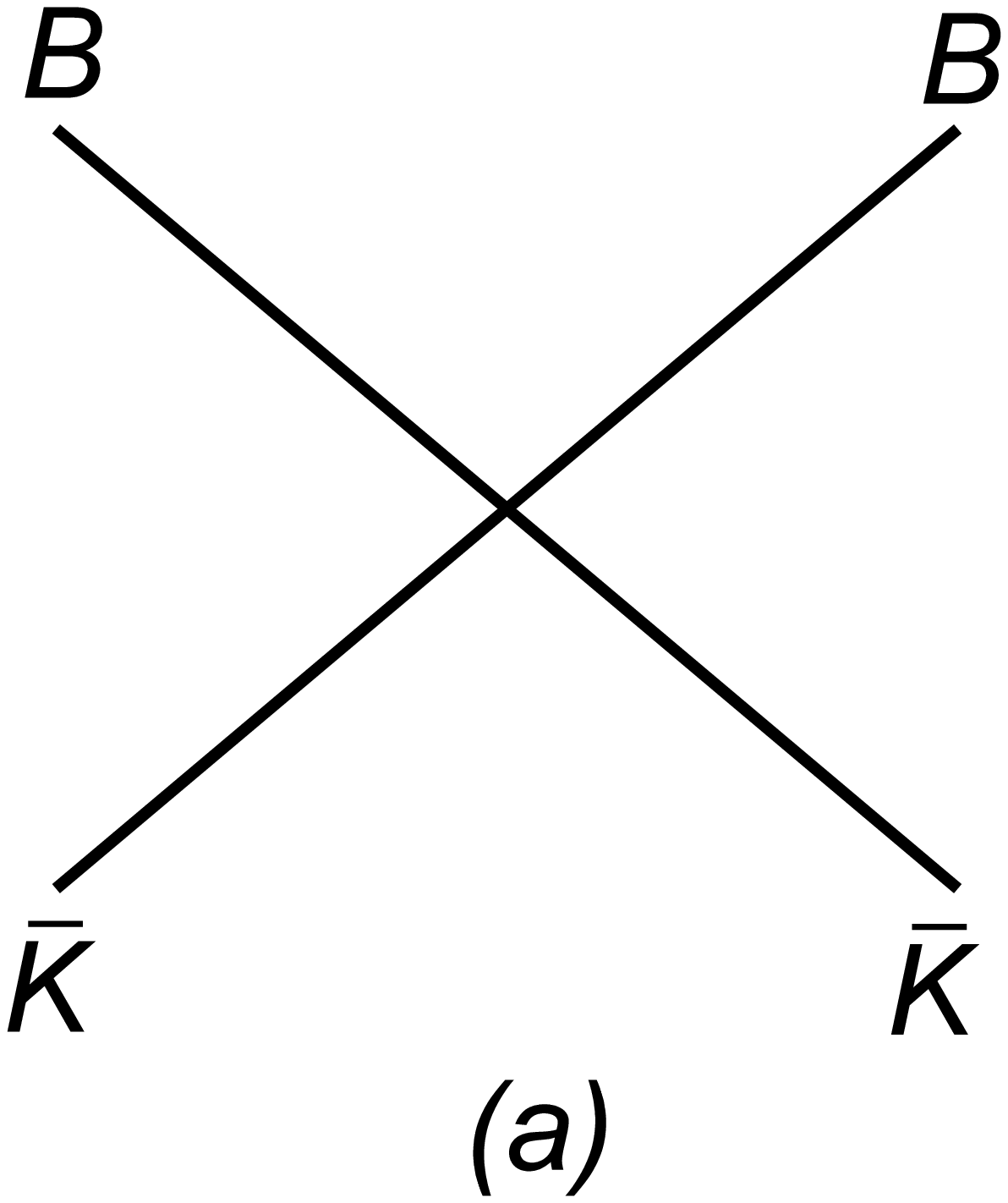}\quad\quad
\includegraphics[scale=0.22]{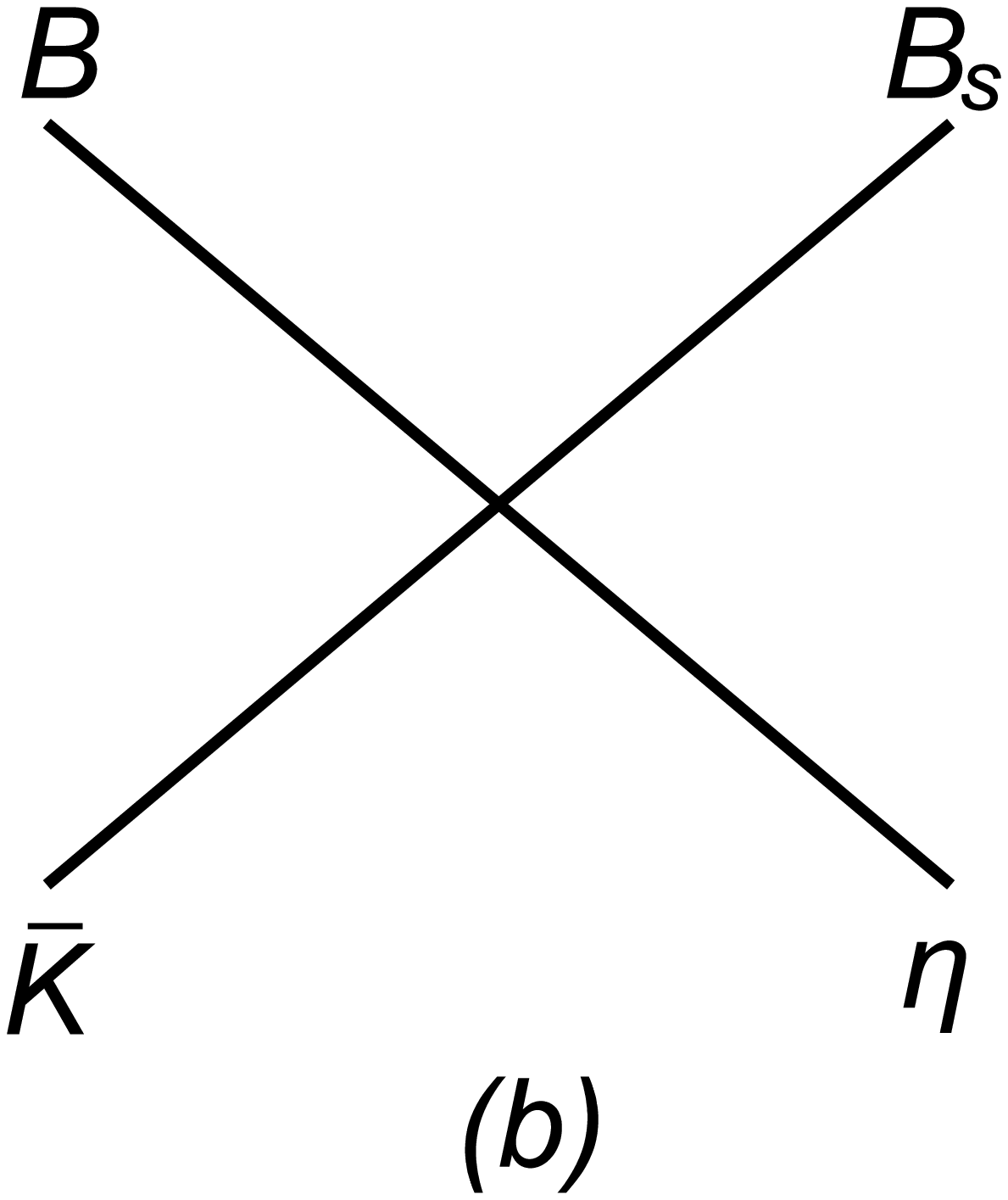}\quad\quad
\includegraphics[scale=0.22]{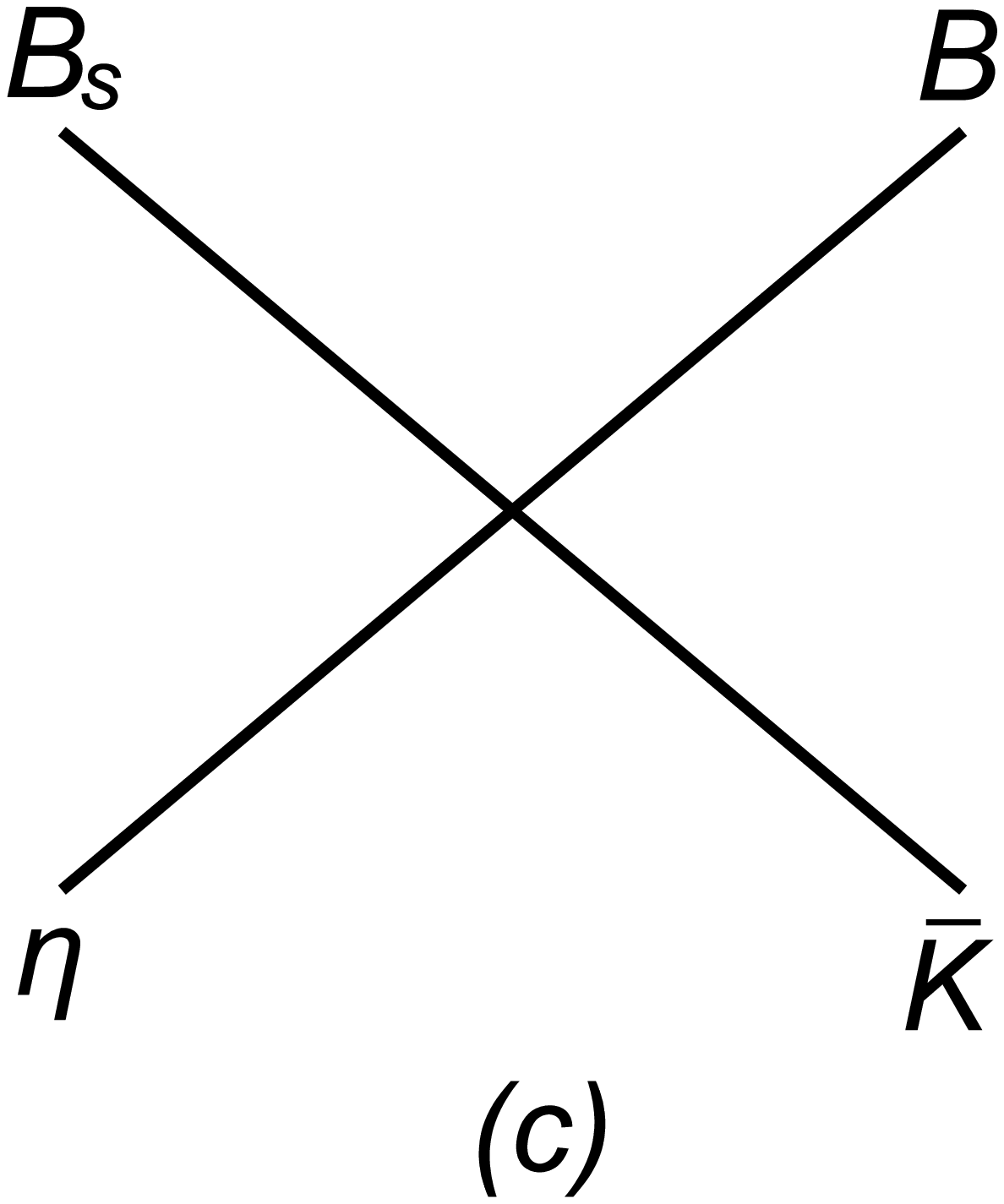}\quad\quad
\includegraphics[scale=0.22]{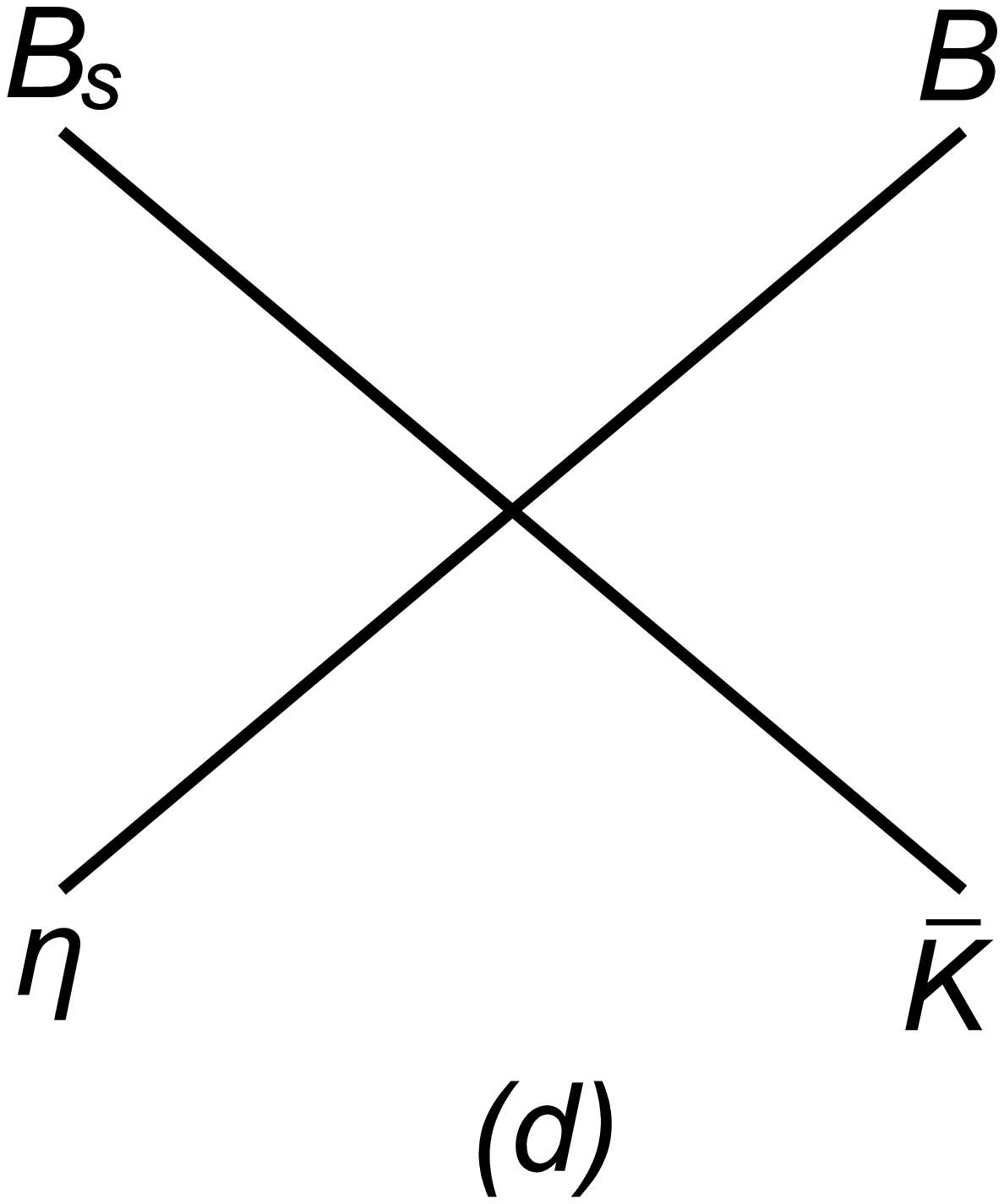}\\
\\
\\
\includegraphics[scale=0.22]{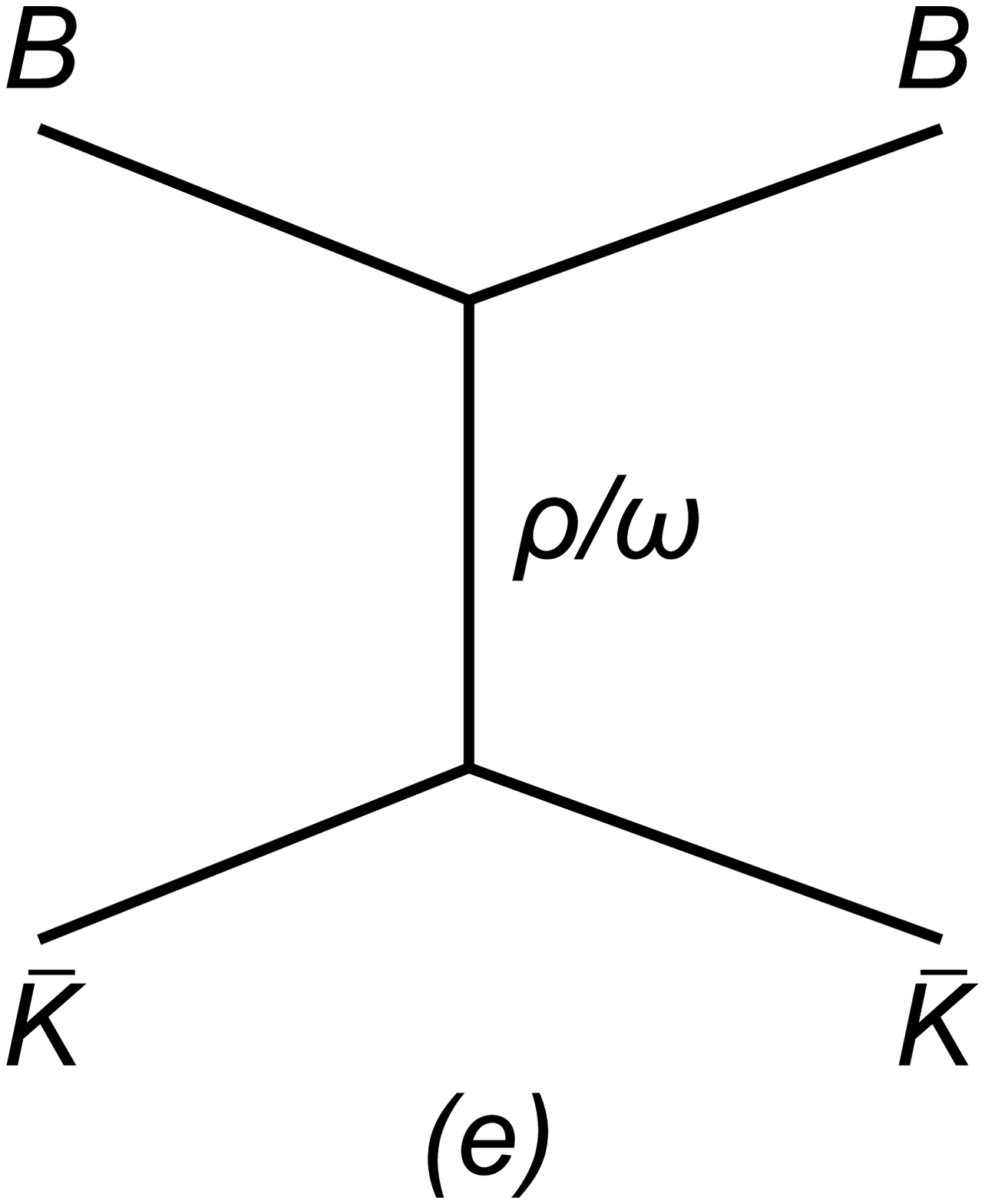}\quad\quad
\includegraphics[scale=0.22]{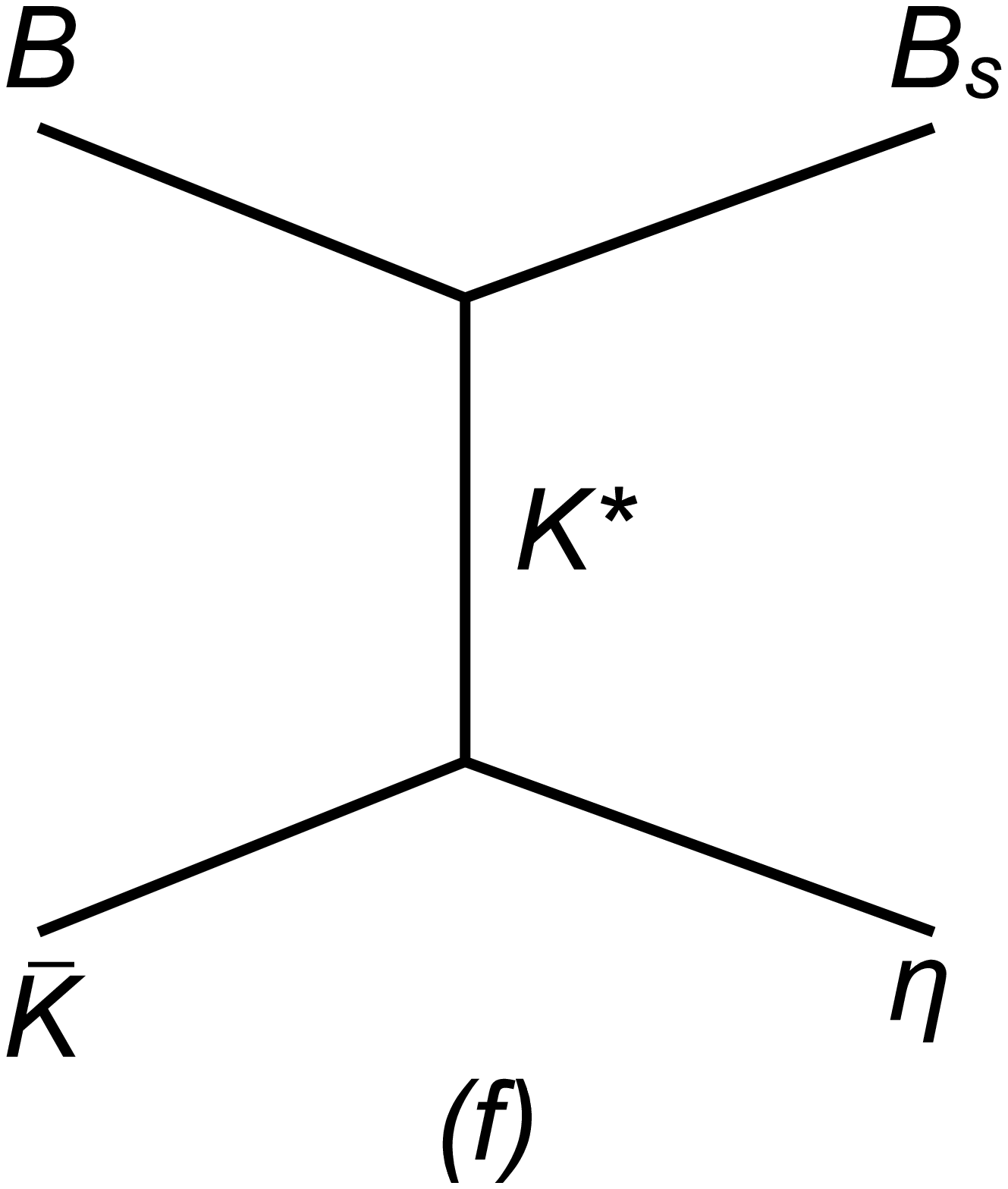}\quad\quad
\includegraphics[scale=0.22]{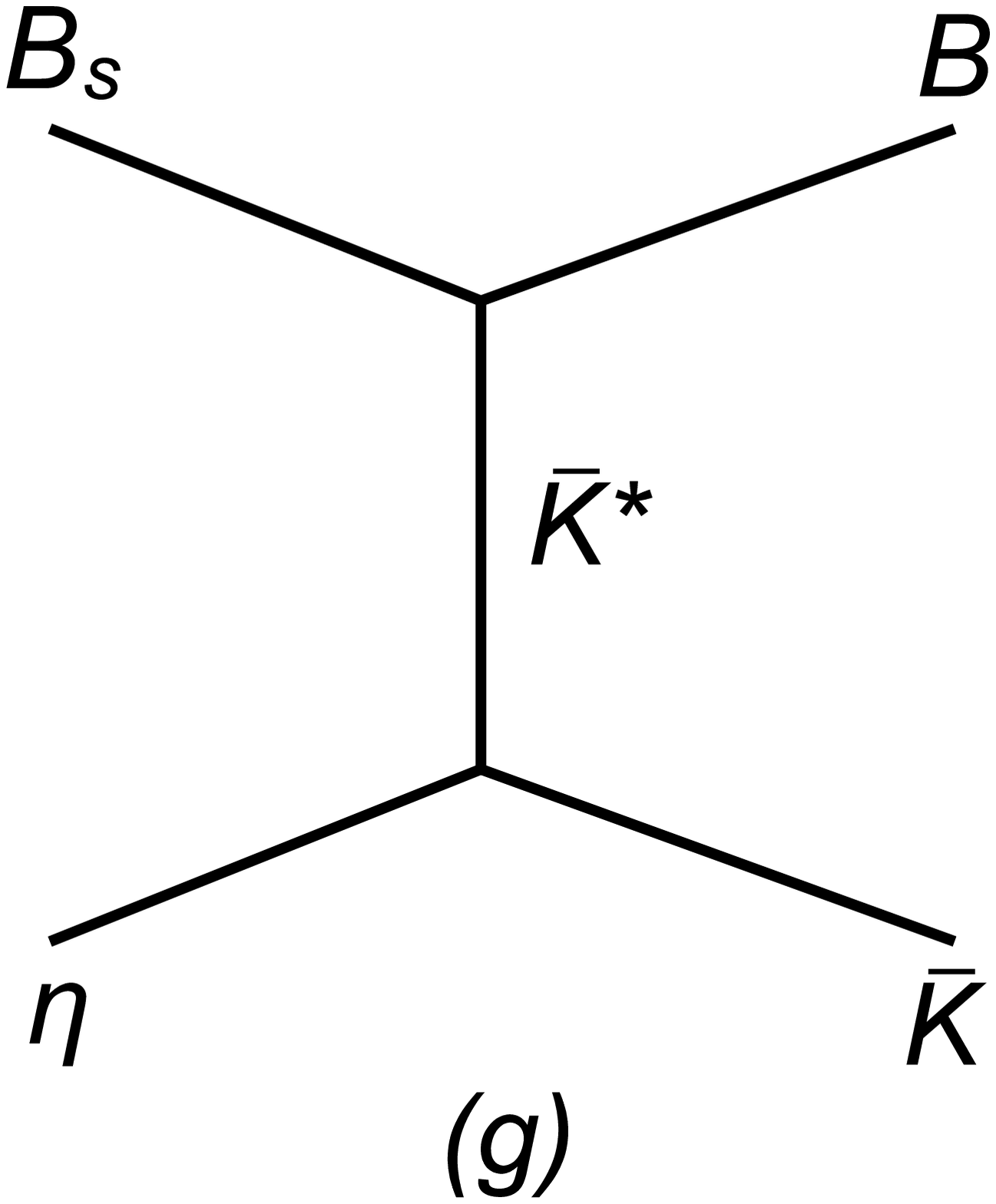}
\end{tabular}
\caption{Feynman diagrams describing $\bar{K}B$ and $\eta B_s$
interaction. \label{fig4}}
\end{figure*}
In Fig. \ref{fig4}, we show the diagrams depicting the interaction
of pseudoscalar and pseudoscalar mesons. The amplitude of contact
terms corresponding to Eq. \eqref{LPPPP} are obtained for
$\bar{K}B\to \bar{K}B$ process in $I=0$ as
\begin{eqnarray}
t_{cont}^{S=0}&=&-\frac{1}{6f^2}(2u-t-s),
\end{eqnarray}
for $\bar{K}B\to \eta B_s$ process
\begin{eqnarray}
t_{cont}^{S=0}&=&-\frac{\sqrt{6}}{12f^2}(s-u)
\end{eqnarray}
and for $\eta B_s\to \eta B_s$ process
\begin{eqnarray}
t_{cont}^{S=0}&=&-\frac{1}{36f^2}(-2t+u+s)
\end{eqnarray}
with $t=(k_1-k_3)^2$. The amplitude of t-channel diagrams for
$\bar{K}B\to \bar{K}B$ have the following expressions
\begin{eqnarray}
t_{ex}^{S=0}&=&-\frac{g^2}{2}\left(\frac{3}{m_\rho^2}+\frac{1}{m_\omega^2}\right)(s-u),
\end{eqnarray}
and for $\bar{K}B\to \eta B_s$
\begin{eqnarray}
t_{ex}^{S=1}&=&-\frac{2\sqrt{6}g^2}{3m_{K^*}^2}(s-u).
\end{eqnarray}
The t channel diagrams for $\eta B_s\to \eta B_s$ has 0
contribution.

\subsection{T-matrix}
With the preparation above, using the Bethe-Salpeter equation in its
on-shell factorized form, we obtain the T-matrix
\begin{eqnarray}
T=(I-VG)^{-1}V,
\end{eqnarray}
where $V$ corresponds to the transition amplitudes shown above, but
projected to $s$-wave. So we neglect the product $\vec{k}_1\cdot
\vec{k}_3$ in the Mandelstam variables $u$ and $t$ which corresponds
to $p$-wave contribution, i.e.,
\begin{eqnarray}
u&\approx& \frac{m_1^2+m_2^2+m_3^2+m_4^2}{2}-\frac{(m_4^2-m_3^2)(m_1^2-m_2^2)}{2s},\nonumber\\
t&\approx&
\frac{m_1^2+m_2^2+m_3^2+m_4^2}{2}+\frac{(m_4^2-m_3^2)(m_1^2-m_2^2)}{2s}.\nonumber\\
\end{eqnarray}
$G$ is the two-meson loop function
\begin{eqnarray}\label{eqG1}
G=i\int
\frac{d^4q}{(2\pi)^4}\frac{1}{q^2-m_1^2+i\epsilon}\frac{1}{(P-q)^2-m_2^2+i\epsilon}.
\end{eqnarray}
Using a cut off of the three momentum, we have
\begin{eqnarray}\label{eqG2}
G=\int_0^{q_{max}}\frac{q^2dq}{(2\pi)^2}\frac{\omega_1+\omega_2}{\omega_1\omega_2[(P^0)^2-(\omega_1+\omega_2)^2+i\epsilon]}.
\end{eqnarray}
This integral was already done (see Ref. \cite{Oller:1998hw}), and
we show it as follows
\begin{eqnarray}\label{eqG3}
G&=&\frac{1}{32\pi^2}\left[\frac{\nu}{s}\left\{
\log\frac{s-\Delta+\nu
\sqrt{1+\frac{m_1^2}{q_{max}^2}}}{-s+\Delta+\nu
\sqrt{1+\frac{m_1^2}{q_{max}^2}}} \right. + \log\frac{s+\Delta+\nu
\sqrt{1+\frac{m_1^2}{q_{max}^2}}}{-s-\Delta+\nu
\sqrt{1+\frac{m_1^2}{q_{max}^2}}}\right\}
-\frac{\Delta}{s}\log\frac{m_1^2}{m_2^2}\nonumber\\
&&+2\frac{\Delta}{s}\log\frac{1+\sqrt{1+\frac{m_1^2}{q_{max}^2}}}{1+\sqrt{1+\frac{m_2^2}{q_{max}^2}}}+\log\frac{m_1^2m_2^2}{q_{max}^2}\left.-2\log\left[\left(1+\sqrt{1+\frac{m_1^2}{q_{max}^2}}\right)\left(1+\sqrt{1+\frac{m_2^2}{q_{max}^2}}\right)\right]\right].
\end{eqnarray}
In Eqs. (\ref{eqG1}), (\ref{eqG2}) and (\ref{eqG3}), $P$ is the
total four-momentum of the two mesons in the loop, $m_1$ and $m_2$
are the masses, $q_{max}$ stands for the cut off,
$\omega_i=\sqrt{\vec{q}_i^2+m_i^2}$, $P^0$ is nothing but the
center-of-mass energy $\sqrt{s}$, $\Delta=m_2^2-m_1^2$, and
$\nu=\sqrt{[s-(m_1+m_2)^2][s-(m_1-m_2)^2]}$.


\section{Results}
\subsection{Discussion of the couplings under SU(4) symmetry}
\begin{figure*}
\begin{tabular}{cccccc}
\includegraphics[scale=0.32]{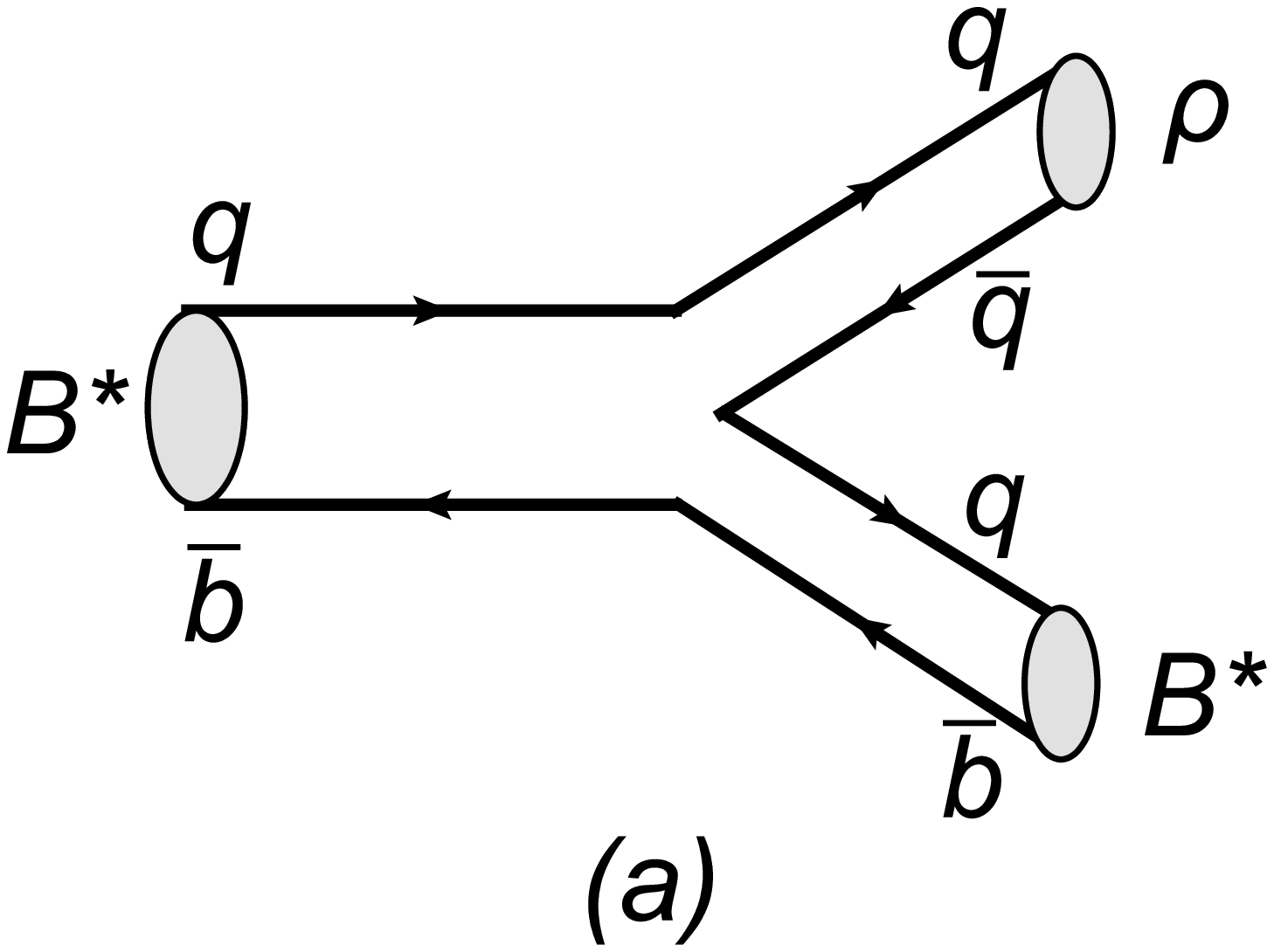}\quad\quad
\includegraphics[scale=0.32]{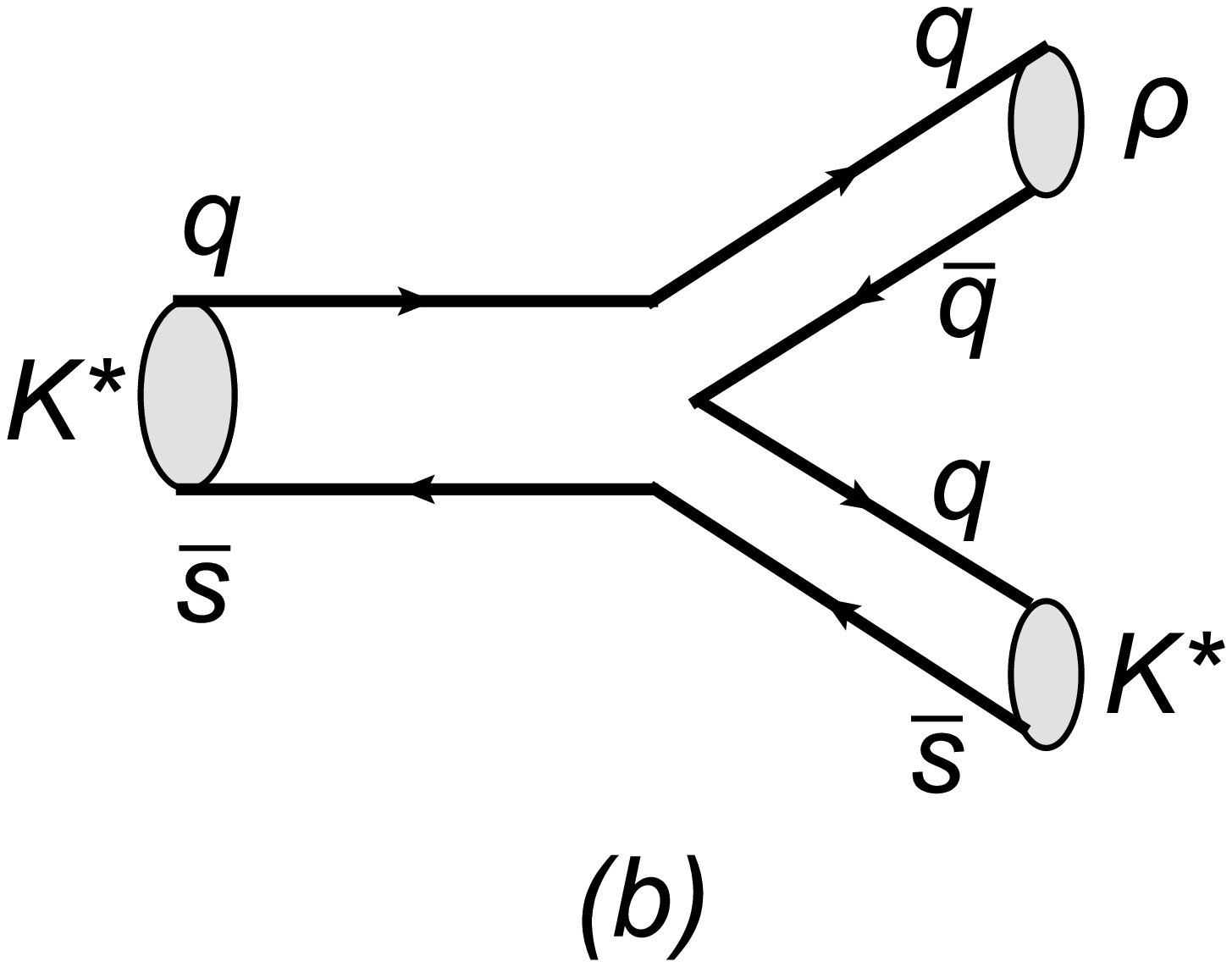}\quad\quad
\end{tabular}
\caption{The vertex of $B^*B^*\rho$ and $K^*K^*\rho$ at the hadronic
level. \label{fig5}}
\end{figure*}
In this subsection, we follow Refs.
\cite{Liang:2014eba,Soler:2015hna,Sakai:2017avl} and discuss the
couplings in the Lagrangian. As an example, we consider the vertex
of $B^*B^*\rho$. In order to estimate the corresponding coupling, we
need to compare this vertex with that of $K^*K^*\rho$, since their
topology is the same if the $\bar{s}$ and $\bar{b}$ quarks are seen
as spectators. Fig. \ref{fig5} shows the diagrams for these two
vertices at the quark level, in which case the corresponding $S$
matrices should be the same, i.e.,
\begin{eqnarray}
S^{mic}&=&1-it\sqrt{\frac{2m_L}{2E_L}}\sqrt{\frac{2m^\prime_L}{2E^\prime_L}}\sqrt{\frac{1}{2\omega_\rho}}
\frac{1}{\mathcal{V}^{3/2}}(2\pi)^4\delta(P_{in}-P_{out}).
\end{eqnarray}
On the other hand, at the hadronic level, the $S$ matrices are
written as
\begin{eqnarray}
S^{mac}_{B^*}&=&1-it_{B^*}\frac{1}{\sqrt{2\omega_{B^*}}}\frac{1}{\sqrt{2\omega_{B^*}}}\frac{1}{\sqrt{2\omega_\rho}}
\frac{1}{\mathcal{V}^{3/2}}(2\pi)^4\delta(P_{in}-P_{out}),\\
S^{mac}_{K^*}&=&1-it_{K^*}\frac{1}{\sqrt{2\omega_{K^*}}}\frac{1}{\sqrt{2\omega_{K^*}}}\frac{1}{\sqrt{2\omega_\rho}}
\frac{1}{\mathcal{V}^{3/2}}(2\pi)^4\delta(P_{in}-P_{out}).
\end{eqnarray}
As discussed above, we should have $S^{mac}_{B^*}=S^{mac}_{K^*}$
which tells us that the corresponding $T$ matrices obey the relation
at the threshold as follows
\begin{eqnarray}\label{Trelation}
\frac{t_{B^*}}{t_{K^*}}=\frac{m_{B^*}}{m_{K^*}}.\label{Trelation}
\end{eqnarray}
If we use the Lagrangian in Eq. (\ref{LVVV}) and calculate the $T$
matrices of the processes in Fig. \ref{fig5}, we find that Eq.
\eqref{Trelation} holds automatically, when the $\rho$ is the
exchanged (virtual) vector meson, because the amplitude has the
$\partial^\mu \cong \partial^0$ operator acting on the external
vectors. The coupling of $B^*B^*\rho$ in Eq. (\ref{LVVV}) implements
correctly the field correction factor of Eq. \eqref{Trelation}.
Since in this case the $b$ quark acts as a spectator in the vertex,
automatically this amplitude is consistent with heavy quark spin
symmetry \cite{manohar}. Similar discussions can be applied to the
$BB\rho$ vertex with respect to $KK\rho$, and we have
\begin{eqnarray}
\frac{t_{B}}{t_{K}}=\frac{m_{B}}{m_{K}},
\end{eqnarray}
but this is what we obtain from Eq. \eqref{LVPP} using SU(4) flavor
symmetry. Effectively one is using SU(3) when the heavy quark is
considered as a spectator. In summary, we apply the Lagrangians of
section II-A, and this takes automatically into account all the
elements discussed above.

\subsection{The $\bar{K}^*B^*$ system}
\begin{figure*}
\begin{tabular}{cccccc}
\includegraphics[scale=0.35]{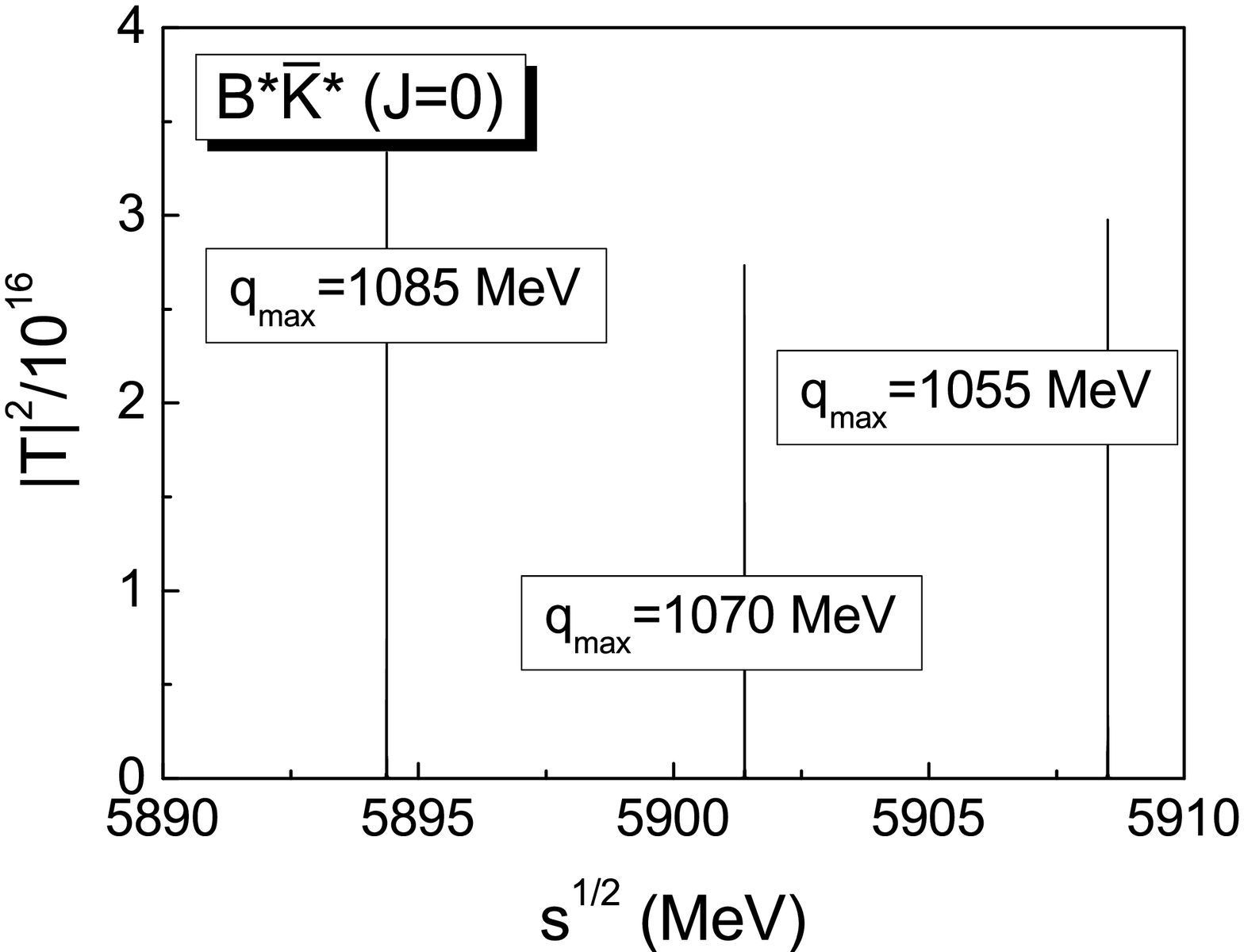}
\includegraphics[scale=0.35]{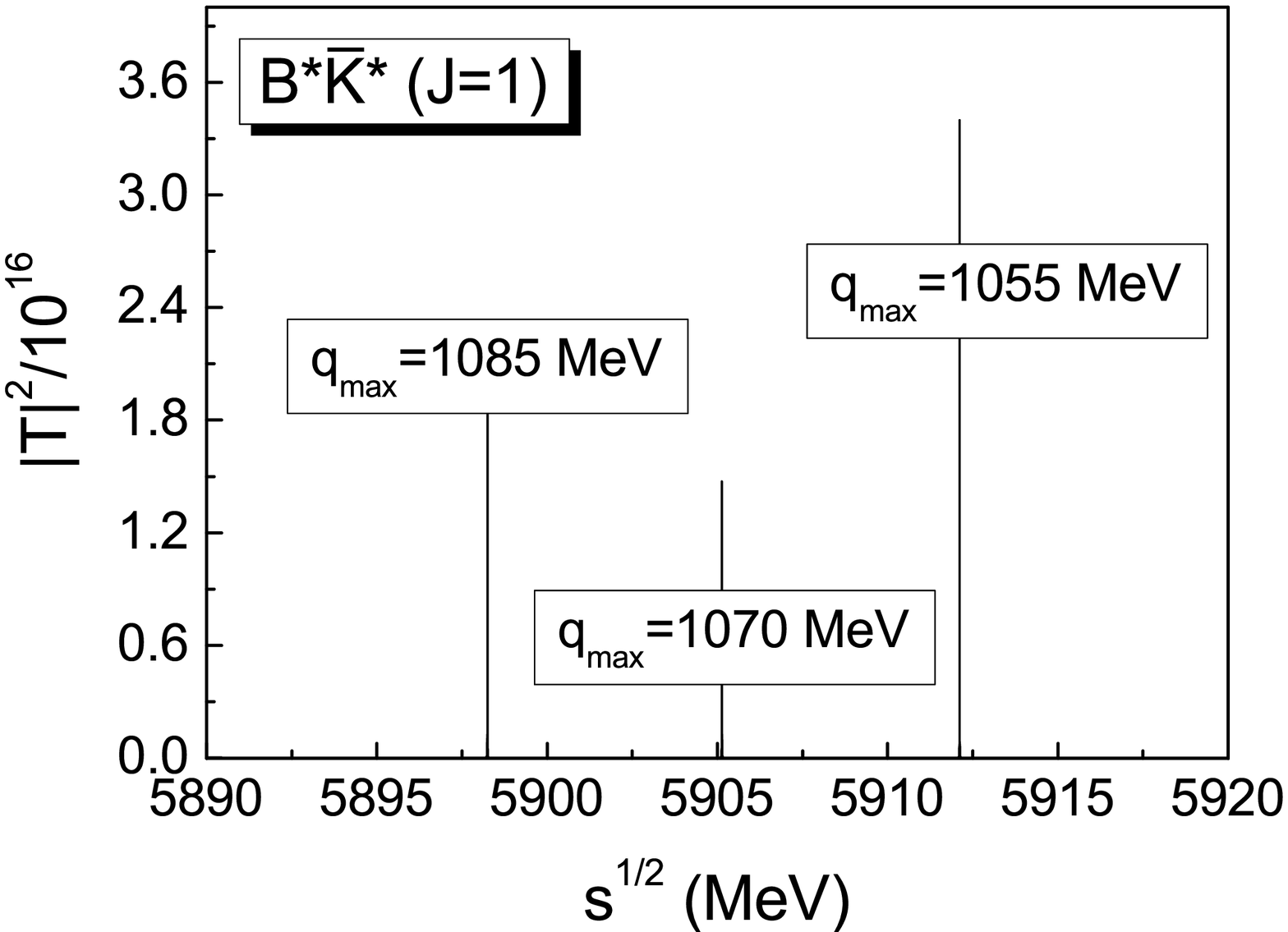}\\
\includegraphics[scale=0.35]{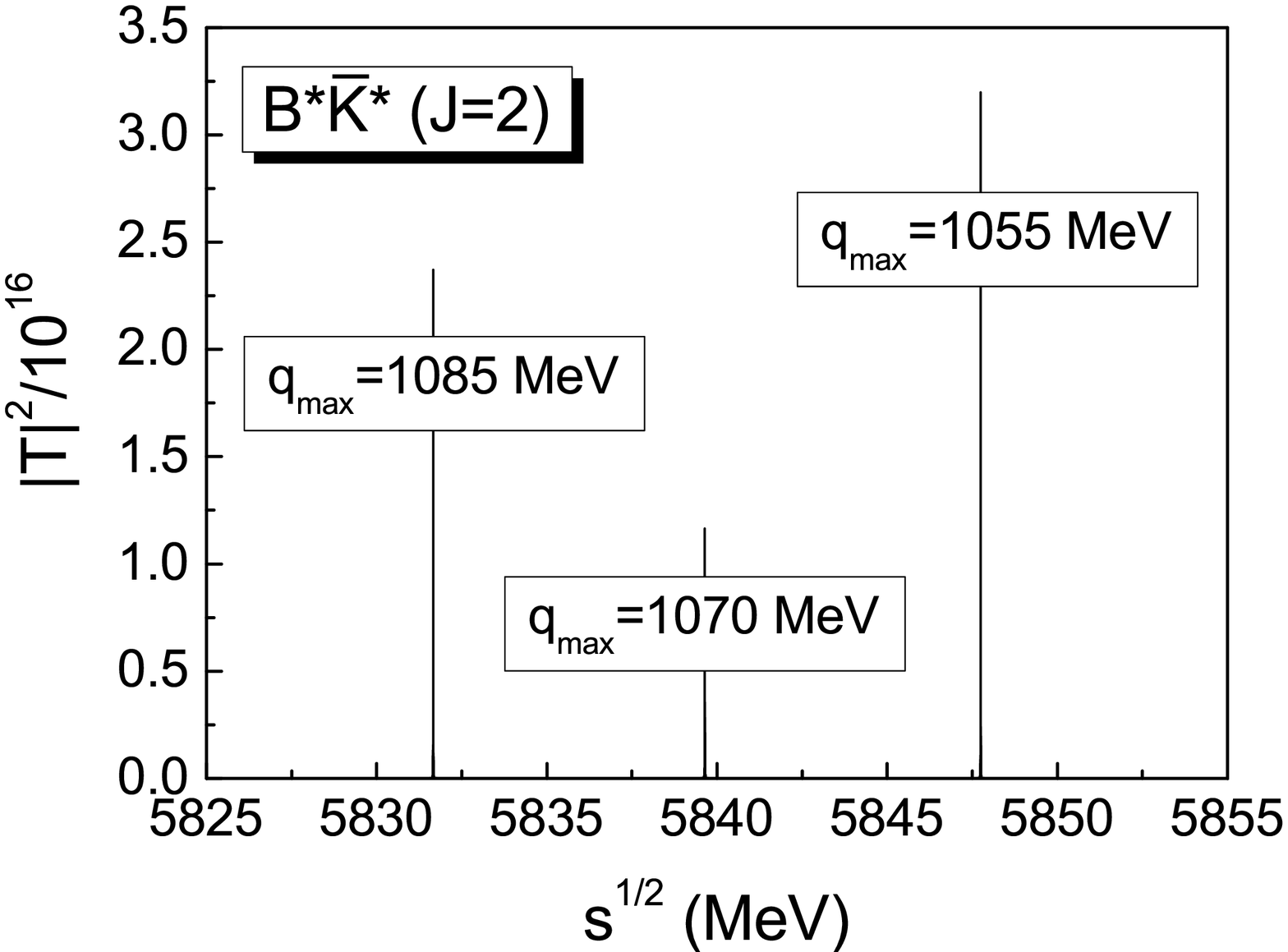}
\end{tabular}
\caption{Squared amplitude for $\bar{K}^*B^*/\omega B_s^*/ \phi
B_s^*$ systems with spin 0, 1 and 2, respectively. \label{fig4.5}}
\end{figure*}

With the potentials given in above section, we solve the
Bethe-Salpeter equation considering $\bar{K}^*B^*$, $\omega B_s^*$
and $\phi B_s^*$ coupled channels. And we obtain three bound states
with $J=0,1,2$, using the cutoff $q_{max}$ around $1055\sim 1085$
MeV. The obtained mass is $5847.8\sim 5831.7$ MeV for the spin 2
state which is consistent with that of $B_{s2}^*(5840)$. With this
$q_{max}$, we predict that the bound state with $J=0$ has a mass
$5908.5\sim 5894.4$ MeV, and the one with $J=1$ has a mass of
$5912.1\sim 5898.2$ MeV. In Fig. \ref{fig4.5}, we plot the line
shape of the mass distribution of these three states. In the PDG
\cite{Patrignani:2016xqp}, the mass of $B_{s1}(5830)$ with spin 1 is
smaller than that of $B_{s2}^*(5840)$. However, the generated bound
state with spin 1 has a mass about $65$ MeV larger than that of the
bound state with spin 2. Henceforth, it is difficult to explain the
$B_{s1}(5830)$ as the $\bar{K}^*B^*$ bound state. In the next
subsection, we will come back to this problem.

The T-matrix close to a pole behaves like
\begin{eqnarray}
T_{ij}&\approx&\frac{g_ig_j}{z-z_R},
\end{eqnarray}
where $i,j=\bar{K}^*B^*,\omega B_s^*, \phi B_s^*$, $g_i$ is the
coupling to the channel $i$, $Re(z_R)$ gives the mass of the bound
state, $Im(z_R)$ the half width, and $z$ is the complex value of the
Mandelstam variable $s$. The coupling for a certain channel is
obtained as
\begin{eqnarray}
g_i^2&=&\lim_{z\to z_R}T_{ii}(z-z_R).
\end{eqnarray}
The sign of the coupling to the $B^*\bar{K}^*$ channel is chosen as
positive, and those for the other channels are then determined by
the following formula
\begin{eqnarray}
\frac{g_i}{g_j}&=&\lim_{z\to z_R}\frac{T_{ii}}{T_{ij}}.
\end{eqnarray}

\begin{table}[htpb]
\caption{The couplings for $\bar{K}^*B^*$ systems mixing with
$\omega B_s^*, \phi B_s^*$ channels. Here we chose the typical value
of the cut off as $1070$ MeV. All the values are given in units of
MeV.\label{tab1}}
\begin{tabular}{ccccc}\toprule [0.4 pt]
channel & J=0 & J=1 & J=2\\
$\bar{K}^*B^*$&45955&45070&49633\\
$\omega B_s^*$ &-10696&-14810&-15017\\
$\phi B_s^*$   &18614&15702&19409 \\\toprule [0.4 pt]
\end{tabular}
\end{table}

The value of the couplings are listed in Tab. \ref{tab1}, from which
we can see that the $\bar{K}^*B^*$ component is dominant for all the
states.


\subsection{The $\bar{K}^*B$ system}
As mentioned in the previous subsection, the $B_{s1}(5830)$ can not
be explained as $\bar{K}^*B^*$ bound state with spin 1, since in PDG
the mass of $B_{s1}(5830)$ is smaller than that of $B_{s2}^*(5840)$,
which is contrary to our results. Now what we do is trying to
explain the $B_{s1}(5830)$ under the $\bar{K}^*B/\omega B_s$ system.

\begin{figure*}
\begin{tabular}{cccccc}
\includegraphics[scale=0.35]{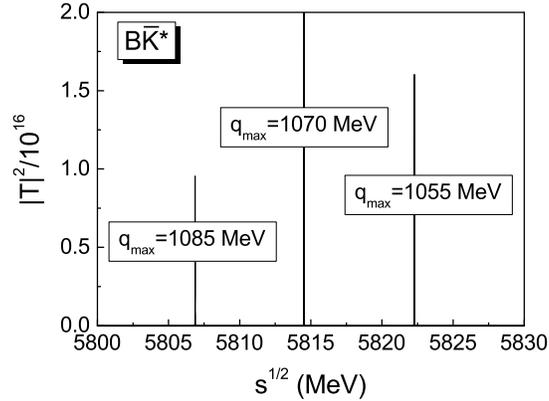}
\end{tabular}
\caption{Squared amplitude for $\bar{K}^*B/\omega B_s/\phi B_s$
sector depending on the center-of-mass energy. \label{fig6}}
\end{figure*}

Under hidden local symmetry there are no contact terms for $VVPP$
vertex, so that only vector exchange diagrams are involved. For the
vector exchange terms, the interactions we study in this subsection
have the same form as that of the $\bar{K}^*B/\omega B_s/\phi B_s$
interactions. So here we expect to find a bound state like in the
case of the $\bar{K}^*B^*$ system. We use $q_{max}=1055\sim 1085$
MeV fixed in the case of $\bar{K}^*B^*$ bound state with spin 2.
Then we obtain a pole position in the range of $5822.3\sim 5806.9$
MeV, which is consistent with the mass of $B_{s1}(5830)$ in the PDG.
In Fig. \ref{fig6}, we plot the line shape of the $|T|^2$ depending
on the center-of-mass energy $\sqrt{s}$. We also calculate the
couplings, which have the value of $g_{\bar{K}^*B}=47654$,
$g_{\omega B_s}=-13388$, $g_{\phi B_s}=18855$ with the cut off
$q_{max}=1070$ MeV.

\subsection{Other predictions}
In this subsection, we will show the results corresponding to
$\bar{K}B^*/\eta B_s^*$ and $\bar{K}B/\eta B_s$ interactions.

\begin{figure*}
\begin{tabular}{cccccc}
\includegraphics[scale=0.35]{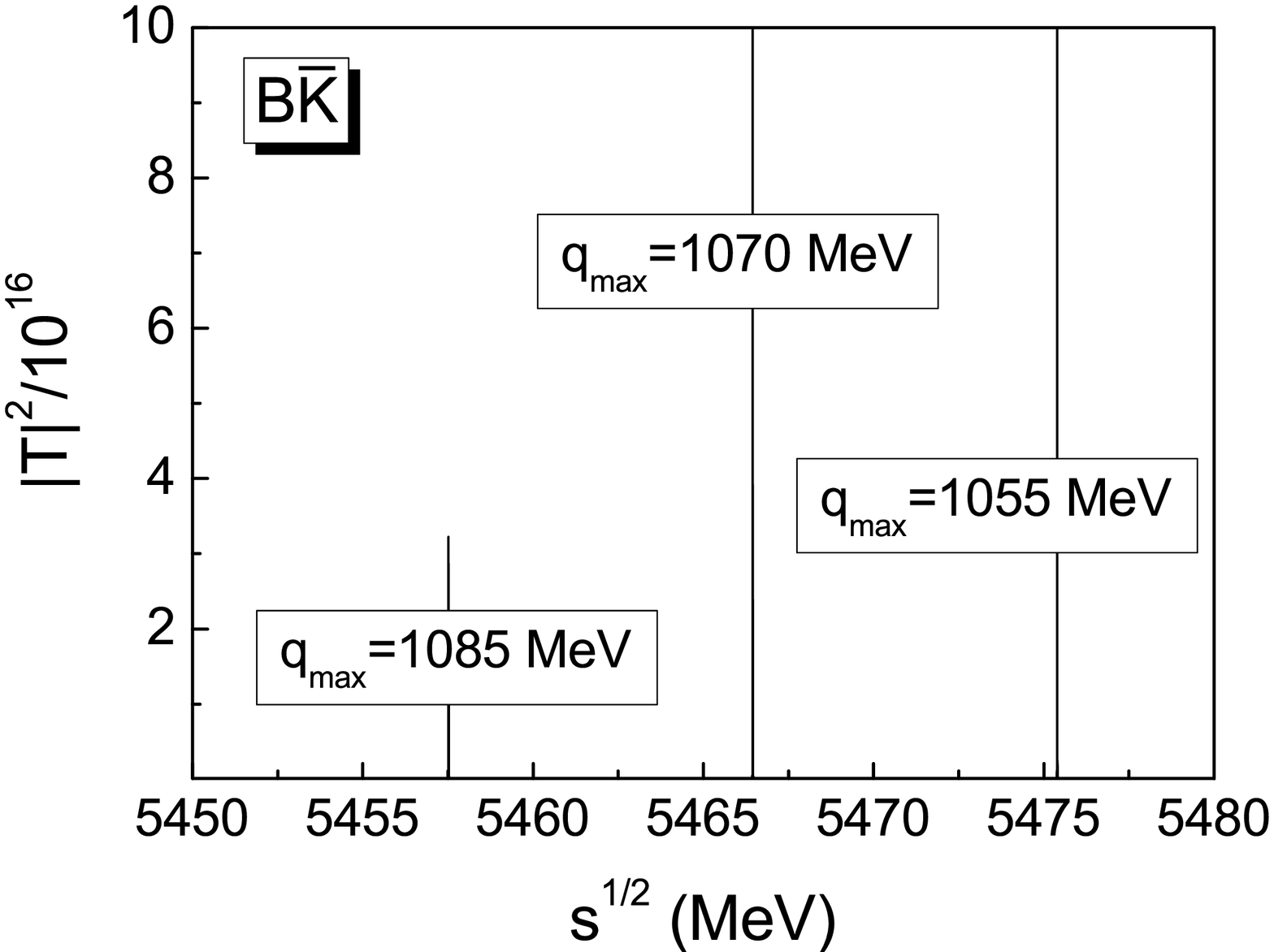}
\includegraphics[scale=0.35]{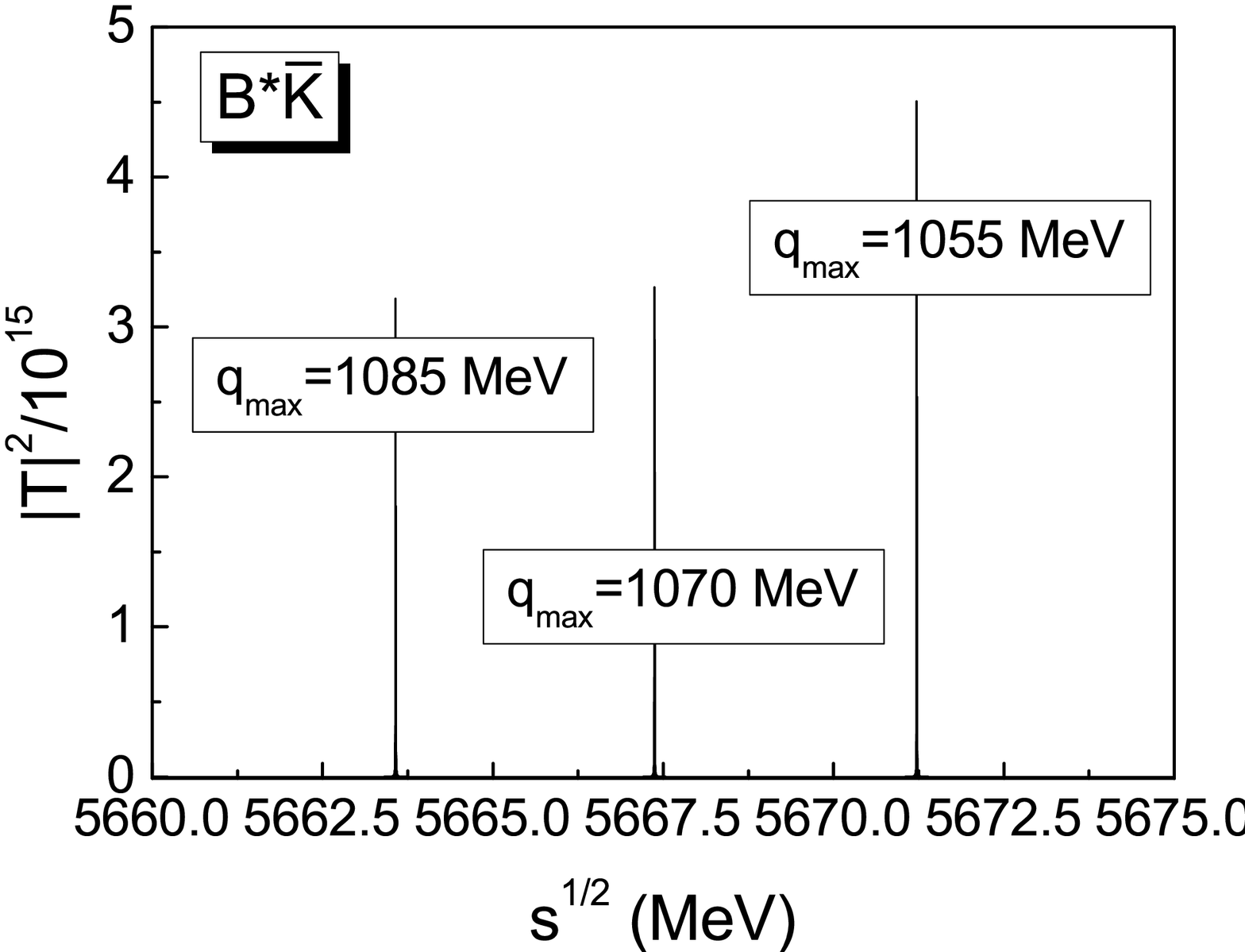}
\end{tabular}
\caption{Squared amplitude for $\bar{K}B/\eta B_s$ and
$\bar{K}B^*/\eta B_s^*$ sector. \label{fig7}}
\end{figure*}

Like the case of $\bar{K}^*B/\omega B_s/\phi B_s$ system, there are
no contact terms for $\bar{K}B^*/\eta B_s^*$ interaction. Only the
vector meson exchange diagrams are considered. In Fig. \ref{fig7},
we plot the squared amplitude depending on the center-of-mass energy
$\sqrt{s}$. Here, we also use the cut off $q_{max}=1055\sim 1085$
MeV as before. The pole position is located at $5671.2\sim 5663.6$
MeV. The couplings of $B^*\bar{K}$ and $B_s^*\eta$ are $30637$ MeV
and $-13919$ MeV respectively, where we choose the cut off as 1070
MeV.

For $\bar{K}B/\eta B_s$ system, we predict a bound state with a mass
of $5475.4\sim 5457.5$ MeV, and the couplings $g_{\bar{K}B}=53577$
MeV and $g_{\eta B_s}=-3689$ MeV, with a cut off $q_{max}=1070$ MeV.

In TABLE. \ref{tab2}, we list our results of all the systems.

\begin{table}
\caption{Summary of our results where the cut off is in the range of
$1055\sim 1085$ MeV and the masses in this table is in the unit of
MeV.\label{tab2}}
\begin{tabular}{c|c|c|c||c|c|c|c}\toprule [0.4 pt]
State mass& $I(J^p)$ & Main component & Exp. &State mass& $I(J^p)$ & Main component & Exp.\\
$5475.4\sim 5457.5$&$0(0^+)$&$\bar{K}B$&-&$5908.5\sim 5894.4$ &$0(0^+)$&$\bar{K}^*B^*$&-\\
$5671.2\sim
5663.6$&$0(1^+)$&$\bar{K}B^*$&-&$5912.1 \sim 5898.2$&$0(1^+)$&$\bar{K}^*B^*$&-\\
$5822.3\sim 5806.9$&$0(1^+)$&$\bar{K}^*B$&$B_{s1}(5830)$&$5847.8\sim
5831.7$&$0(2^+)$&$\bar{K}^*B^*$&$B_{s2}^*(5840)$\\\toprule [0.4 pt]
\end{tabular}
\end{table}

\section{summary}
In this work, we have studied the systems containing bottomed and
strange quarks by the chiral unitary approach. Considering
$\bar{K}^*B^*$ and $\omega B_s^*$ coupled channels and solving the
Bethe-Salpeter equation, we find three states with masses
$5908.5\sim 5894.4$ MeV, $5912.1\sim 5898.2$ MeV and $5847.8\sim
5831.7$ MeV, with the cut off $q_{max}$ chosen as $1055\sim 1085$
MeV. The state with spin 2 can be identified with $B_{s2}^*(5840)$.
From the couplings that we obtained, we can see that the
$\bar{K}^*B^*$ component is dominant. However, the $B_{s1}(5830)$
can not be explained as the state with spin 1, since its mass is
smaller than that of $B_{s2}^*(5840)$. So we studied another system,
i.e., $\bar{K}^*B/\omega B_s$ system, and we get a bound state with
a mass $5822.3\sim 5806.9$ MeV which agrees with the mass of
$B_{s1}(5830)$. In addition, we also studied $\bar{K}B^*/\eta B_s^*$
and $\bar{K}B/\eta B_s$ interactions, and predict two bound states
with masses $5671.2\sim 5663.6$ MeV and $5475.4\sim 5457.5$ MeV,
respectively. We expect further experiments to confirm our
predictions.

\vfil

\section*{Acknowledgments}
This work is partly supported by the National Science Foundation for
Young Scientists of China under Grants NO. 11705069 and the
Fundamental Research Funds for the Central Universities. It is
partly supported by the National Natural Science Foundation of China
(Grants No. 11475227, 11735003) and the Youth Innovation Promotion
Association CAS (No. 2016367). This work is also partly supported by
the Spanish Ministerio de Economia y Competitividad and European
FEDER funds under the contract number FIS2011-28853-C02-01, FIS2011-
28853-C02-02, FIS2014-57026-REDT, FIS2014-51948-C2- 1-P, and
FIS2014-51948-C2-2-P, and the Generalitat Valenciana in the program
Prometeo II-2014/068.


\end{document}